\documentclass[fleqn,usenatbib]{mnras}

\usepackage{newtxtext,newtxmath}

\usepackage[T1]{fontenc}
\usepackage{ae,aecompl}

\usepackage{graphicx}	
\usepackage{amsmath}	
\usepackage{amssymb}	
\usepackage{subfig}     
\usepackage{threeparttable} 

\title[Polarized spectral index between 30--44\,GHz]{The spectral index of polarized diffuse Galactic emission between 30 and 44\,GHz}

\author[L. Jew et al.]{
Luke Jew,$^{1}$\thanks{E-mail: luke.jew@physics.ox.ac.uk}
R.D.P. Grumitt,$^{1}$
\\
$^{1}$Department of Physics, University of Oxford, Denys Wilkinson Building, Keble Road, Oxford, OX1 3RH\\
}

\date{Accepted XXX. Received YYY; in original form ZZZ}

\pubyear{2015}

\begin{document}
\label{firstpage}
\pagerange{\pageref{firstpage}--\pageref{lastpage}}
\maketitle

\begin{abstract}
We present an estimate of the polarized spectral index between the \textit{Planck} 30 and 44\,GHz surveys in $3.7^\circ$ pixels across the entire sky.
We use an objective reference prior that maximises the impact of the data on the posterior and multiply this by a maximum entropy prior that includes information from observations in total intensity by assuming a polarization fraction.
Our parametrization of the problem allows the reference prior to be easily determined and also provides a natural method of including prior information. 
The spectral index map is consistent with those found by others between surveys at similar frequencies.
Across the entire sky we find an average temperature spectral index of $-2.99\pm0.03(\pm1.12)$ where the first error term is the statistical uncertainty on the mean and the second error term (in parentheses) is the extra intrinsic scatter in the data.
We use a clustering algorithm to identify pixels with actual detections of the spectral index.
The average spectral index in these pixels is $-3.12\pm0.03(\pm0.64)$ and then when also excluding pixels within $10^\circ$ of the Galactic plane we find $−2.92(\pm0.03)$.
We find a statistically significant difference between the average spectral indices in the North and South \textit{Fermi} bubbles.
Only including pixels identified by the clustering algorithm, 
the average spectral index in 
the southern bubble is $-3.00\pm0.05(\pm0.35)$, which is similar to the average across the whole sky.
In the northern bubble we find a much harder average spectral index of $-2.36\pm0.09(\pm0.63)$.
Therefore, if the bubbles are features in microwave polarization they are not symmetric about the Galactic plane.
\end{abstract}




\section{Introduction}
Diffuse Galactic synchrotron radiation is emitted by cosmic ray electrons and positrons spiralling through the Galactic magnetic field \citep{Rybicki1985}.
In polarization, this radiation is the dominant foreground to the cosmic microwave background (CMB) below around 100\,GHz \citep{Page2007}.
Unlike the already detected CMB temperature fluctuations or CMB $E$-mode signal, the primordial $B$-mode signal is possibly weaker than the polarized synchrotron foreground at all frequencies,  angular scales and areas of the sky \citep{Dunkley2009}.
Better constraints on the frequency spectrum of diffuse Galactic synchrotron radiation are therefore vital if the primordial $B$-mode signal is to be unambiguously detected.

The frequency spectrum of synchrotron radiation is determined by the energy distribution of the radiating cosmic rays,
which is approximately a power law above a few GeV for around two decades in energy \citep{Casadei2004,Aguilar2014}.
Diffuse Galactic synchrotron radiation therefore has a power-law frequency spectrum over approximately four decades in frequency.

Except in regions with continuous injection of new relativistic cosmic-ray electrons, the synchrotron spectrum has a cut-off at high frequencies caused by radiative energy loss.
At low-frequency ($\lesssim100\,\mathrm{MHz}$) the spectrum turns over due to self absorption \citep{Condon2016}.
Deviations from a power law can be introduced by localised variations in the distribution of cosmic-ray energies, by line-of-sight effects \citep{Chluba2017}, and in polarization by Faraday rotation \citep[for example][]{Lazarian2016}.

For diffuse Galactic synchrotron radiation, observations from tens of megahertz to hundreds of gigahertz  find an average total-intensity temperature spectral index between 2.5-3.5  \citep[for example][]{Lawson1987,Reich1988,Platania2003,Davies2006,Guzman2011,PlanckCollaboration2015,Ade2016}.
These measurements agree with simulations such as \citet{Orlando2013}.

Synchrotron radiation can be highly polarized, in principle up to 75 per cent \citep{Rybicki1985}.
At high Galactic latitude diffuse Galactic synchrotron is measured to be up to $\sim40$ per cent polarized but is more typically 20 per cent polarized with much lower polarization fractions close to the Galactic plane \citep[for example][]{Kogut2007,Vidal2014,Ade2016}.

A summary of measurements of the average spectral index of polarized diffuse Galactic synchrotron radiation is presented in Table~\ref{tab:real_polBetas}.
The tabulated results were obtained using a variety of methods and datasets.
The methods include template fitting, $T-T$ plots and parametric fitting. 
The datasets include those from space-based CMB satellites \textit{WMAP} \citep{Bennett2013,Jarosik2011,Hinshaw2009,Bennett2003a} and  \textit{Planck} \citep{PlanckCollaboration2018b,PlanckCollaboration2016a}, and ground-based surveys like S-PASS \citep{Carretti2019}, C-BASS \citep{Jones2018} and PGMS \citep{Carretti2010}.
Many of the analyses listed in Table~\ref{tab:real_polBetas} assumed a single spectral index across the sky.

\begin{table*}
    \centering
    \caption{Estimates of the spectral index of polarized diffuse Galactic synchrotron emission at frequencies above $\sim5\,\mathrm{GHz}$.}
    \label{tab:real_polBetas}
    \begin{threeparttable}
    \begin{tabular}{rrrlll}
        \hline\hline
        Frequency & Geometric mean & Spectral index & Notes &Comments & Reference \\
        range [GHz] & frequency [GHz] & $\beta$& & &\\
        \hline
        All \textit{WMAP} & - & $-3.2$ & & All-sky &\citet{Kogut2007}\\
        ---\texttt{"}--- & - & $-2.98\pm0.01$ &$^a$ & Galactic plane & ---\texttt{"}---\\
        ---\texttt{"}--- & - & $-3.12\pm0.04$ &$^a$ & High latitude &---\texttt{"}---\\
        
        ---\texttt{"}--- & - & $-3.03\pm0.04$ &$^a$ & Where $\sigma(\beta)<0.25$ &\citet{Dunkley2008}\\
        ---\texttt{"}--- & - & $-3.00\pm0.04$ &$^a$ & $|b|<20^\circ$ &---\texttt{"}---\\
        
        ---\texttt{"}--- & - & $-3.08\pm0.06$ &$^a$ & $|b|>20^\circ$ &---\texttt{"}---\\
        
        All \textit{Planck} & - & $-3.5$ -- $-3.0$ &$^b$ & All-sky, \textsc{Commander2} &\citet{PlanckCollaboration2018a}\\
        
        ---\texttt{"}--- & - & $-3.1\pm0.06$ &$^a$ & All-sky, \textsc{SMICA} &---\texttt{"}---\\
        
        1.4--22.8&5.6& $-3.21\pm0.15 $ &$^c$ &$|b|>30^\circ$ & \citet{Carretti2010}\\
        
        2.3--33 & 8.7& $-3.25\pm0.15$ &$^c$ & $\delta<-1^\circ$, $|b|>20^\circ$&  \citet{Krachmalnicoff2018}\\
        
        4.78--28.4& 11.7&$-3.1658\pm0.0002$ &$^a$ &$\delta>-12^\circ$ &\citet{Jew2017}\\
        
        22.5--32.6 &27.1 & $-2.99\pm0.01$ &$^a$ & All-sky &\citet{Fuskeland2014}\\
        ---\texttt{"}--- &---\texttt{"}---& $-2.98\pm0.01$ &$^a$ & Galactic plane &---\texttt{"}---\\
        ---\texttt{"}--- &---\texttt{"}--- & $-3.12\pm0.04$ &$^a$ & High latitude &---\texttt{"}---\\
        
        23, 33 and 41 & 31.5 & $-3.06\pm0.02$ & $^a$ & Average over 18 regions & \citet{Vidal2014}\\
        
        28.4--44.1 & 35.4& $-2.99(\pm0.03)\pm1.12$ &$^d$ & Whole sky & This work (Section~\ref{sec:realResults})\\
        ---\texttt{"}--- & ---\texttt{"}---& $-3.12(\pm0.03)\pm0.64$ &$^d$ & Clustered points & ---\texttt{"}---\\
        ---\texttt{"}--- & ---\texttt{"}---& $-2.92(\pm0.03)\pm0.48$ &$^d$ & Clustered points, $|b|>10^\circ$ &---\texttt{"}--- \\
        
        \hline
    \end{tabular}
    \begin{tablenotes}
    \item $^a$ Error term only includes the statistical uncertainty on the mean and no intrinsic scatter of the data. When calculated over large numbers of pixels these uncertainties become very small.
    \item $^b$ The marginalized posterior peaks in this range.
    \item $^c$ Error term includes intrinsic scatter of data and uncertainty on mean.
    \item $^d$ First error term (in parentheses) is the uncertainty on the mean, the second error term is the intrinsic scatter of the data.
    \end{tablenotes}
    \end{threeparttable}
\end{table*}

Variations in the spectral index typically trace structures in our Galaxy.
At radio frequencies the most prominent structures are the radio loops and spurs, the most pronounced of which is the North Polar Spur (Loop I) \citep{Roger1999}.
These loops are most likely caused either by the shock-waves from supernova explosions accelerating electrons in the inter-stellar medium or stellar winds from OB associations \citep{Berkhuijsen1971,Heiles1980,Salter1983,Wolleben2007,Mertsch2013,Vidal2014}.
In the shock waves of recent supernova explosions the high energy electrons have not had time to radiate away their energy and this can result in hard (more shallow) spectra,
older populations of cosmic-ray leptons emit softer (steeper) synchrotron emission.

One particularly interesting feature is the \textit{WMAP} Haze.
The Haze is a diffuse structure with a total intensity spectral index of about $\beta\simeq-2.5$ that is centred on the Galactic Centre \citep{Finkbeiner2004,Dobler2008,Dobler2012a,PlanckCollaboration2012a}.
The Haze is not symmetric about the Galactic plane, it is weaker in the south \citep{Su2010},
and this north/south asymmetry could be caused by differences in the circumgalactic medium \citep{Sarkar2018}.
It is important to note that the Haze has many overlapping features including the Gould Belt (particularly north of the plane), and it spans a large region covering many tens of kiloparsecs including the Galactic centre.

On the sky, the extent of the \textit{WMAP} Haze approximately coincides with that of the \textit{Fermi} Bubbles, a double-lobed feature observed in Fermi-LAT data that extends to Galactic latitudes of $b=\pm55^\circ$ \citep{Dobler2010,Su2010,Dobler2012}.
Unlike the Haze, the \textit{Fermi} Bubbles are approximately symmetric about the Galactic plane and have 
well defined edges.
A number of mechanisms for producing both the \textit{Fermi} Bubbles and the \textit{WMAP} Haze have been proposed including annihilation of Dark Matter particles \citep{Finkbeiner2004a} and AGN-like activity at the Galactic centre \citep{Su2012,Yang2013}.

Detection of the haze in polarization has proven more difficult, \citet{Gold2011} found no indication for a haze of polarized hard spectrum synchrotron radiation.
Polarized outflows from the Galactic centre have been observed that spatially correlate with the bubbles/haze \citep{Jones2012,Carretti2013a} and could be related \citep{Crocker2014}.

In this work we use Bayesian methods to estimate the spectral index from maps of polarized intensity. 
Our parametrization of the problem allows the objective reference prior for the problem to be easily determined and information from total intensity observations to be easily included.
Much of the publicly available low-frequency polarization surveys (here defined to be below 50\,GHz) are either low signal-to-noise or corrupted by Faraday rotation.
We use the \textit{Planck} 30\, and 44\,GHz surveys
as
these are two of the highest signal-to-noise polarized all-sky maps publicly available, that are also not so low in frequency for Faraday effects to be significant (except along the Galactic plane).
We do not use the \textit{WMAP} polarization maps as they have poorly constrained large-scale modes \citep{Jarosik2011}.
These modes effectively introduce varying zero-levels across the $Q$ and $U$ maps that corrupt the polarized intensities.

We make a careful choice of prior that maximises the impact of the data on the posterior and that also includes weakly informative prior information from total intensity data.
We use the relative entropy between the posterior and the prior along with a clustering algorithm to identify pixels with good detections of the spectral index.

The paper is laid out as follows.
In Section~\ref{sec:model_and_method} we introduce our method and justify our choice of prior and posterior distributions.
In Section~\ref{sec:data} we explain the processing that we applied to the data used in this work.
In Section~\ref{sec:simsResults} we demonstrate that the method works well on simulated data.
In Section~\ref{sec:realResults} we present our results using the method on real data and show that our results are generally consistent with the work of others.
We find that the North and South \textit{Fermi} bubbles have different average spectral indices. 
The spectral index in the northern bubble is shallower than in the southern bubble and we discuss possible causes for this observation.
We conclude in Section~\ref{sec:conclusions}.

\section{Model and method} \label{sec:model_and_method}
We want to estimate the spectral index from two noisy measurements of the polarized intensity at different frequencies, and we also want to incorporate our prior knowledge into the analysis.
Instead of working with maps of the Stokes $Q$ and $U$ parameters, we use the polarized intensities.
The polarized intensity is insensitive to Faraday rotation (although not depolarization) and also allows us to more easily encode our prior knowledge from total intensity data.
Calculating the analytic form of the posterior distribution of the spectral index is beyond the scope of this work and so instead we estimate it numerically.

We do this by sampling from the posterior probability distribution of the polarized intensities, which is constructed by the multiplication of a likelihood function by a prior \citep[and a normalizing constant,][]{Bayes1763},
\begin{equation}
    p(A_1,A_2|P_1,P_2) \propto p(P_1,P_2|A_1,A_2)p(A_1,A_2)
\end{equation}
where $A_i$ and $P_i$ are the true and measured polarized intensities at frequency $\nu_i$ respectively.

The spectral index between polarized emission at two different frequencies, $\nu_1$ and $\nu_2$, in the $n^\textrm{th}$ pixel of a map is then
\begin{equation}
\label{eqn:beta}
    \beta(n) = \frac{\log\left(A_1(n)/A_2(n)\right)}{\log(\nu_1/\nu_2)}.
\end{equation}

In Section~\ref{sec:likelihood} we describe how
we approximate the polarized intensities as being Rician random variables.
In Section~\ref{sec:prior} we describe how 
our prior is formed by multiplying the objective reference prior by the maximum entropy distribution for the testable information that we wish to include.
In Section~\ref{sec:posterior} we explain how
we explore the resulting posterior distribution for the spectral index with MCMC methods and how we compare the posteriors to the prior.

\subsection{Likelihood function}\label{sec:likelihood}
The polarized intensity, $P$, is formed from the quadrature sum of the measured Stokes $Q$ and $U$ values, which can themselves be modelled as Gaussian random variables.
If the errors on $Q$ and $U$ are equal then the measured values of $P$ are drawn from a Rician distribution 
\begin{align}
    p(P_i|A_i,\sigma_P) &= \frac{P_i}{\sigma_P^2}\exp{\left(-\frac{P_i^2+A_i^2}{2\sigma_P^2}\right)} I_0\left(\frac{P_i A_i}{\sigma_P^2}\right)\\
    & = \textrm{Rician}(P_i,A_i,\sigma_P) \nonumber,
\end{align}
where $I_0$ is the zeroth order modified Bessel function of the first kind and $\sigma_P$ is the standard deviation of noise on $Q$ and $U$.
The likelihood function for the true intensities in a pixel $n$ in maps at frequencies $\nu_1$ and $\nu_2$ is then the product of two Ricians,
\begin{equation}
    \label{eqn:likelihood}
    p(P_1,P_2 \vert A_1, A_2)=
    \prod_{i=1}^2\textrm{Rician}(P_i,A_i,\sigma_{Pi}).
\end{equation}

If the errors on $Q$ and $U$ are not equal then $P$ is distributed according to a Beckmann distribution \citep[also known as a generalized Rician distribution,][]{Beckmann1959,YongjunXie2000,Zhu2018},
\begin{align}
    \label{eqn:beckmann}
    p(P_i&\vert \mu_{Qi},\mu_{Ui}\sigma_{Qi},\sigma_{Ui}) = \frac{P_i}{2\pi\sigma_{Qi}\sigma_{U_i}}\times \nonumber\\ 
    &\int_0^{2\pi}\exp\left[-\frac{(P_i\cos{\theta}-\mu_{Qi})}{2\sigma_{Qi}}
    -\frac{(P_i\sin{\theta}-\mu_{Ui})}{2\sigma_{Ui}}\right]
    \mathrm{d}\theta.
\end{align}
The posterior distributions of the true polarized intensities, and therefore spectral indices, could then be determined from the posterior distributions of $\mu_{Qi}$ and $\mu_{Ui}$.
There are two main reasons that we do not adopt this likelihood function or parametrization.
Firstly, the integral in Equation~\ref{eqn:beckmann} can not be solved analytically and so this likelihood function is expensive to compute.
Secondly, to construct a weakly informative prior in this instance becomes difficult.
The prior must allow for negative values of $\mu_Q$ and $\mu_U$ but there is no maximum entropy distribution for a variable with known expectation value that extends over the entire real line (Section~\ref{sec:prior}) 
and furthermore 
it requires prior information on the expected polarization angles across the sky.

\subsection{Prior} \label{sec:prior}
Away from the Galactic plane and brightest radio loops, the \textit{Planck} polarization data are low signal-to-noise and as such the prior is important.
We want as non-informative a prior on the true polarized intensities as possible, whilst still including weakly informative constraints from other observations.
Our prior is formed by multiplying the objective reference prior for the problem by the maximum entropy prior for the information we want to include.

The reference prior is the prior distribution that maximizes the Kullback-Leibler divergence \citep[or relative entropy,][]{Kullback1951} between itself and the posterior distribution, i.e. it allows the data to have maximal impact on the posterior distribution \citep{Bernardo1979,Bernardo1981,Berger1992,Bernardo2005}.
Reference priors have several properties in their favour over other proposed non-informative priors; the process to generate them is completely general, they are invariant to parameter transformations, and they have consistent marginalization and sampling properties.
They have been used in cosmology by \citet{Heavens2018} to ensure that prior assumptions are not dominating estimates of the neutrino mass hierarchy.

Because the likelihood function in Equation~\ref{eqn:likelihood} can be neatly factorized into the form $p(\vec{P}\vert \vec{A})=\prod_i p(P_i\vert A_i)$, the reference prior is simply the product of the one dimensional Jeffreys priors for each term in the product.
The Jeffreys prior for the Rician disitrbution is $p(A_i)\propto A_i$, which corresponds to a uniform distribution in the $Q-U$ plane \citep{Jeffreys1946,Lauwers2009}.
The reference prior for this problem is therefore
\begin{equation}
    p_\mathrm{ref.}(A_1,A_2) \propto A_1 A_2.
    \label{eqn:rice_ref_prior}
\end{equation}

We are not totally ignorant of the expected polarized intensity of diffuse Galactic emission from 1--100\,GHz and therefore should include this information in our prior.
As stated earlier: synchrotron radiation is typically $\sim20\,\textrm{per cent}$ polarized, its frequency spectrum approximately obeys a power law, and low-frequency total intensity surveys are dominated by synchrotron radiation.
To calculate our prior expected polarized intensity as a function of frequency
we extrapolate the 408\,MHz  Haslam survey (\citet{Haslam1982} as reprocessed by \citet{Remazeilles2014}) to higher frequencies with temperature spectral index of -3.11 and assume a uniform polarization fraction of 0.2 across the sky.
The prior expected polarized intensity could be improved in future work using a spatially varying total intensity spectral index and polarization fraction \citep[for exmaple those produced by][note that strictly the prior should not be formed from data included in the likelihood]{Miville-Deschenes2008}.

We do not have strong prior knowledge on the expected spread of the polarized intensities from this value.
We therefore multiply the reference prior by the maximum entropy prior for a positive valued variable with known expectation value; i.e. an exponential distribution,
\begin{equation}
    p_\mathrm{max.\,ent.}(A_1,A_2) = \prod_{i=1}^2 \lambda_i\exp{(-\lambda_i A_i)}.
    \label{eqn:max_entrop_prior}
\end{equation}

The total prior on parameters $A_1$ and $A_2$ is therefore the product of Equations~\ref{eqn:rice_ref_prior} and \ref{eqn:max_entrop_prior},
\begin{equation}
    \label{eqn:prior}
    p(A_1,A_2) = \prod_{i=1}^2 A_i\lambda_i^2\exp{(-\lambda_i A_i)},
\end{equation}
where $1/\lambda_i$ is the prior expected value of $A_i$ and it is raised to the second power to ensure the prior integrates to unity.

Both the likelihood and prior for $A_i$ are unimodal distributions with positive skew.
The prior is only more peaked than the likelihood when $1/\lambda_i\lesssim\sigma_{Pi}$.
When this condition is met then the peak of the posterior is shifted from that of the likelihood to much closer to the prior.
When it is not met the effect of the prior on the posterior is small.

From the prior distribution in Equation~\ref{eqn:prior}, we can analytically derive the prior distribution on the spectral index $\beta$,
\begin{equation}
    \label{eqn:priorBeta}
    p(\beta) = -\frac{6\lambda_1^2\lambda_2^2 \exp{(2\beta/r)}}{r(\lambda_1\exp{(\beta/r)}+\lambda_2)^4},
\end{equation}
where $r=1/\log{(\nu_1/\nu_2)}$.

\subsection{Posterior distribution of the spectral index parameter} \label{sec:posterior}
A calculation of the analytic form of the posterior distribution for $\beta$ is beyond the scope of this work.
Instead we sample the posterior using Markov Chain Monte Carlo (MCMC) methods.

The posterior distribution for $A_1$ and $A_2$ is the product of Equations~\ref{eqn:likelihood} and \ref{eqn:prior}.
We explore the posterior distribution for $A_1$ and $A_2$ in each pixel using the Metropolis-Hastings algorithm \citep{Metropolis1953,Hastings1970}, implemented using \textsc{PyMC} \citep{Patil2010}.
For each step in the converged chain we calculate the spectral index using Equation~\ref{eqn:beta}, these are then samples drawn from the spectral index posterior distribution.

We can quantify the increase in information on the spectral index in each pixel by calculating the Kullback-Liebler divergence, $D_\mathrm{KL}$, between the analytic equation for the prior (Equation~\ref{eqn:priorBeta}) and our numerical estimate of the posterior distribution \citep{Kullback1951},
\begin{equation}
    D_\mathrm{KL}(p_\mathrm{posterior}||p_\mathrm{prior}) = \int_{-\infty}^{\infty} p_\mathrm{posterior}(\beta) \log\left( \frac{p_\mathrm{posterior}(\beta)}{p_\mathrm{prior}(\beta)}\right) \mathrm{d}\beta.
\end{equation}

\section{Data} \label{sec:data}
In this section we describe the {\it Planck} 30 and 44\,GHz polarized intensity maps \citep{PlanckCollaboration2018c} that we estimate the spectral index from along with the ancillary Haslam total intensity survey, which we use to form a weakly informative prior (Section~\ref{sec:prior}), and the SMICA CMB map \citep{PlanckCollaboration2018a}, which we use to subtract an estimate of the CMB from the \textit{Planck} maps.
The \textit{Planck} maps used in this analysis are summarised in Table~\ref{tab:data_surveys}.
In this work all maps use the \textsc{HEALPix} pixelization scheme \citep{Gorski2005}.

In each pixel, if the noise on $Q$ and $U$ is Gaussian with $\sigma_Q=\sigma_U$ then the polarized intensity, $P=\sqrt{Q^2+U^2}$, is distributed like a Rician random variable.
In the \textit{Planck} 30\,GHz map the Median ratio of $\sigma_Q$ to $\sigma_U$ is $\mathrm{Median}(\sigma_Q/\sigma_U)=1.00$.
In the \textit{Planck} 44\,GHz map $\mathrm{Median}(\sigma_Q/\sigma_U)=1.07$ and therefore $\sigma_Q\not\approx\sigma_U$.
To ensure that $P$ is Rician distributed, we add small amounts of extra noise to the $Q$ and $U$ maps.
In each pixel, if $\sigma_Q>\sigma_U$ then we add a number drawn from a Gaussian distribution with zero mean and variance $\sigma_Q^2-\sigma_U^2$, and vice versa for pixels with $\sigma_U>\sigma_Q$.
The $\sigma_P$ map is then the maximum of $\sigma_Q$ and $\sigma_U$ in each pixel.

\begin{table}
    \centering
    \caption{Summary statistics of the \textit{Planck} surveys used in this work.
    The centre frequency is for emission with a temperature spectral index of $\beta=0.0$.
    The colour corrections are for emission with a temperature spectral index of $\beta=-3.0$.
    The noise refers to the noise levels in maps smoothed to 220\,arcmin and downgraded to \textsc{HEALPix} $N_\mathrm{side}=16$ pixels.}
    \label{tab:data_surveys}
    \begin{tabular}{lrr}
    \hline\hline
         & \textit{Planck} 30\,GHz & \textit{Planck} 44\,GHz\\
    \hline
    Centre frequency [GHz]& 28.4 & 44.1\\
    Colour correction & 0.999& 0.987\\
    $\mathrm{K}_\mathrm{RJ}/\mathrm{K}_\mathrm{CMB}$ & 0.979& 0.951\\
    Median $P$ [$\mu$K$_\mathrm{RJ}$]&7.0&2.2  \\
    Median $\sigma_P$ [$\mu$K$_\mathrm{RJ}$] & 0.6& 0.7\\
    \hline
    \end{tabular}
\end{table}

\begin{figure*}
    \centering
    \subfloat{\includegraphics[width=0.8\columnwidth]{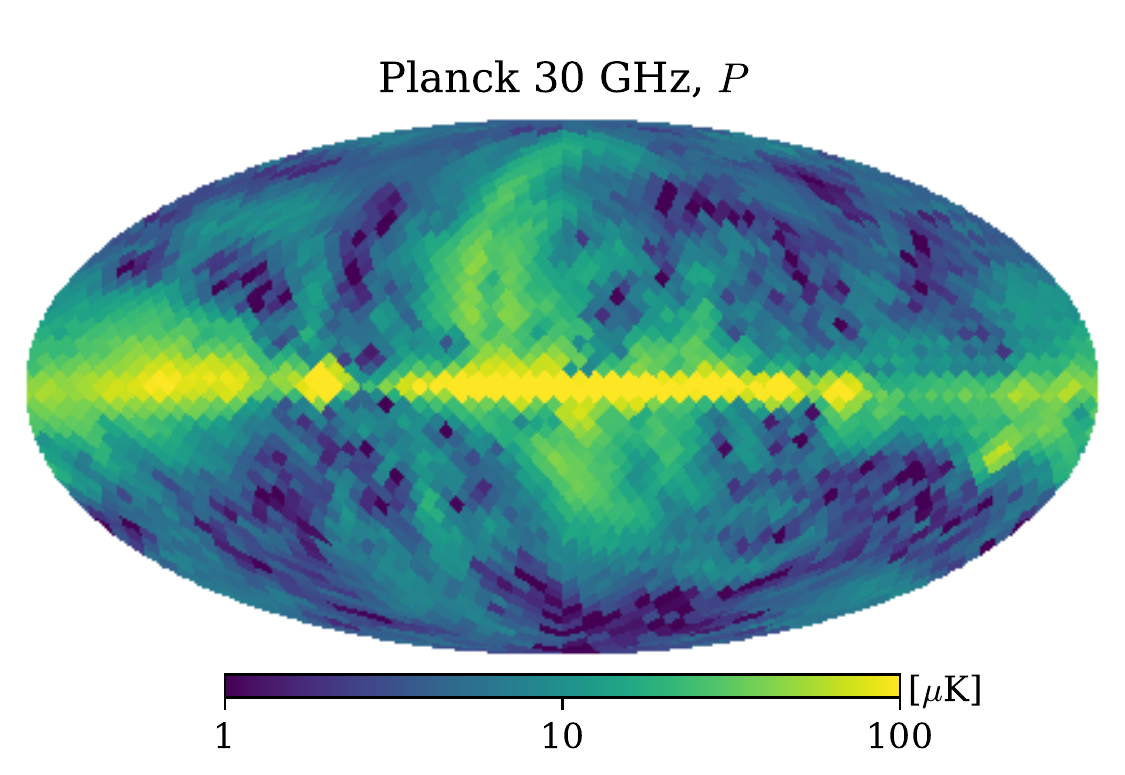}}
    \subfloat{\includegraphics[width=0.8\columnwidth]{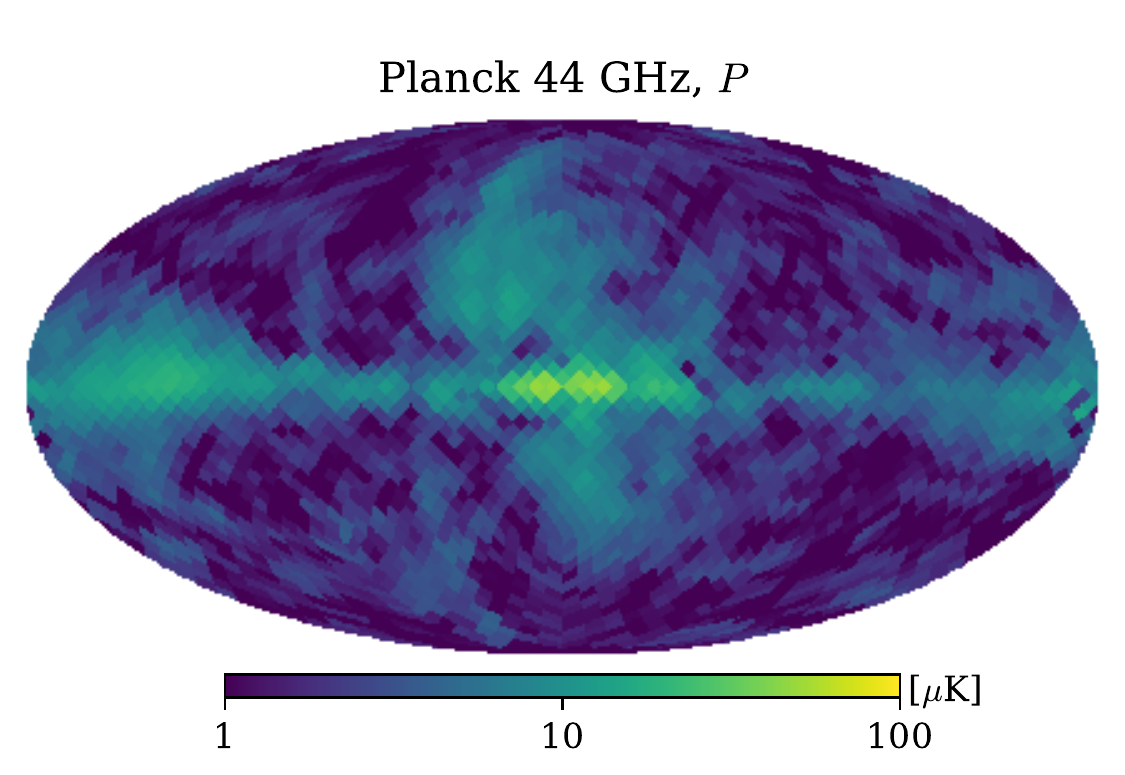}}\\
    \subfloat{\includegraphics[width=0.8\columnwidth]{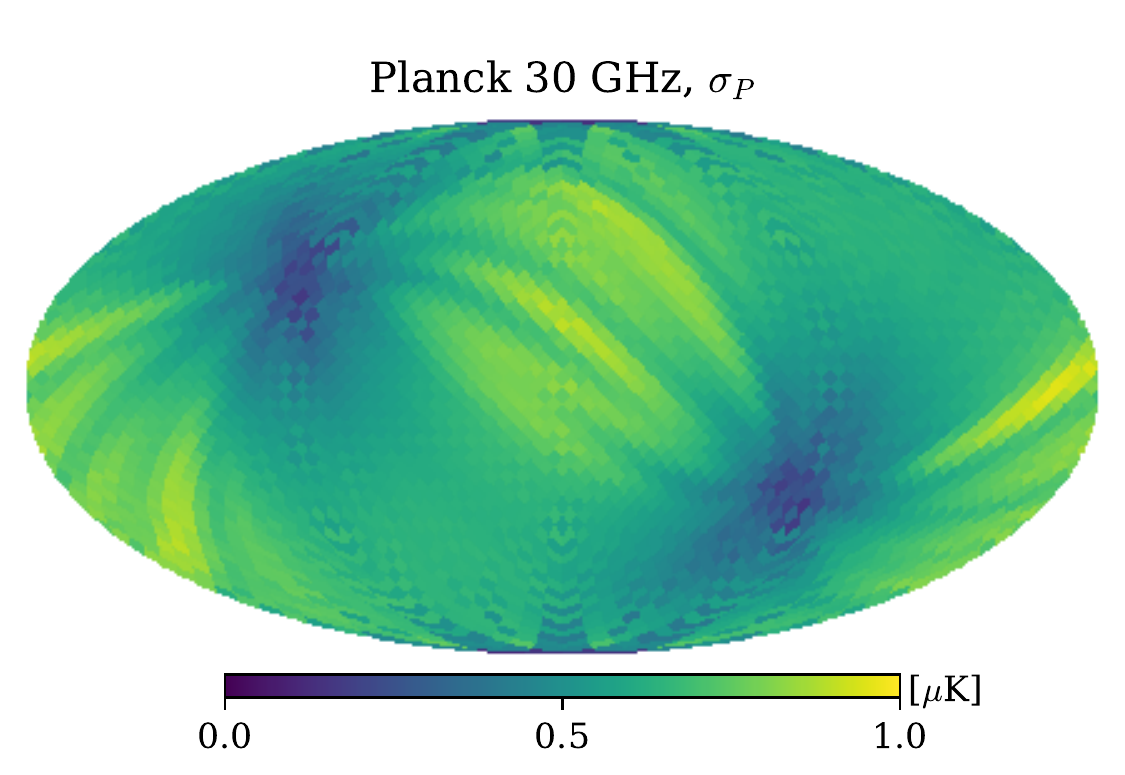}}
    \subfloat{\includegraphics[width=0.8\columnwidth]{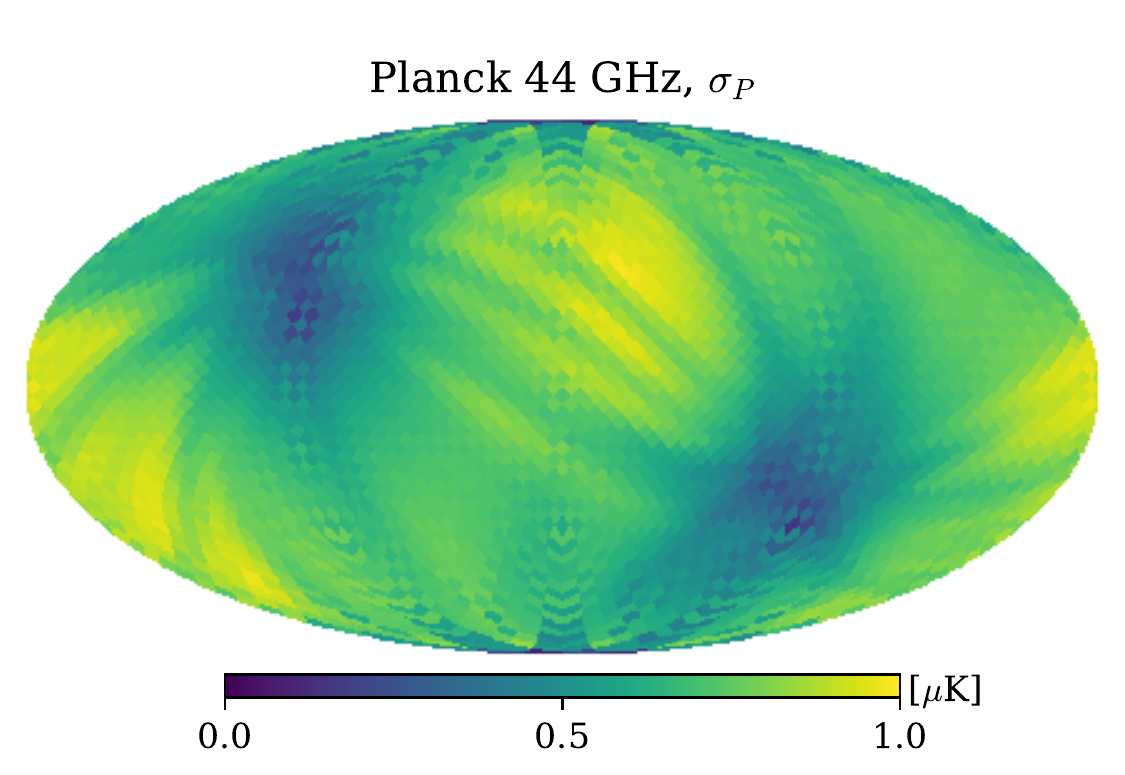}}\\
    \subfloat{\includegraphics[width=0.8\columnwidth]{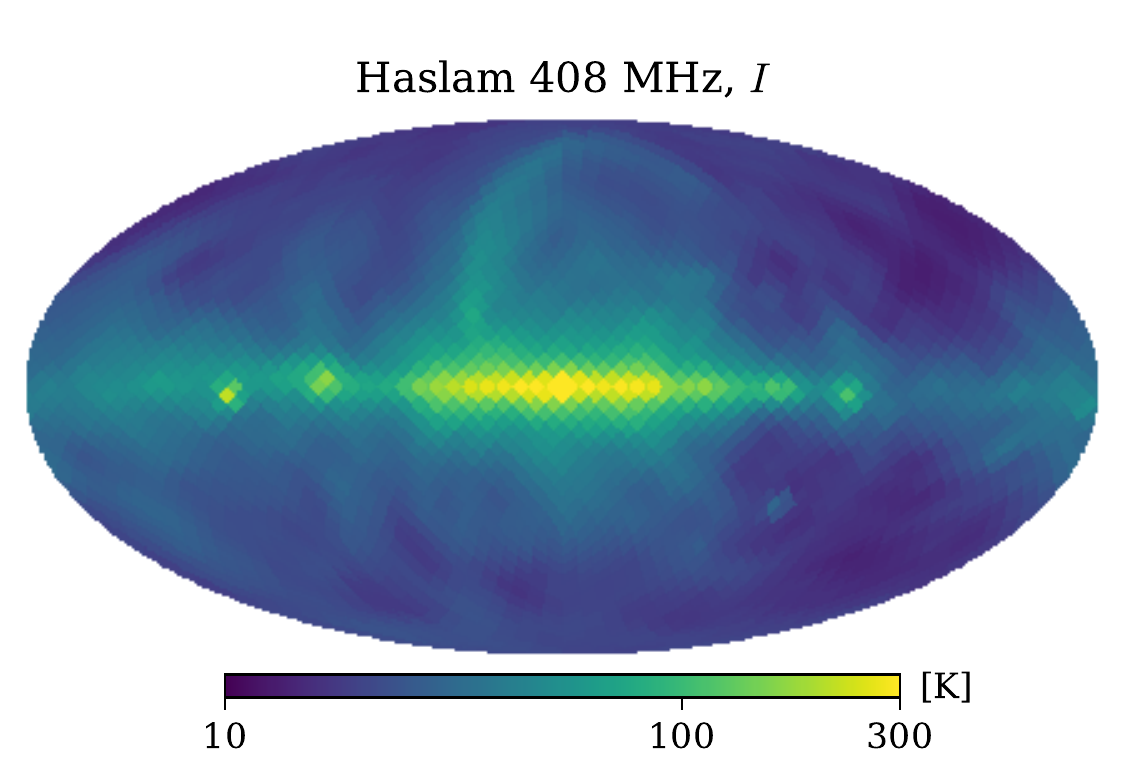}}
    \subfloat{\includegraphics[width=0.8\columnwidth]{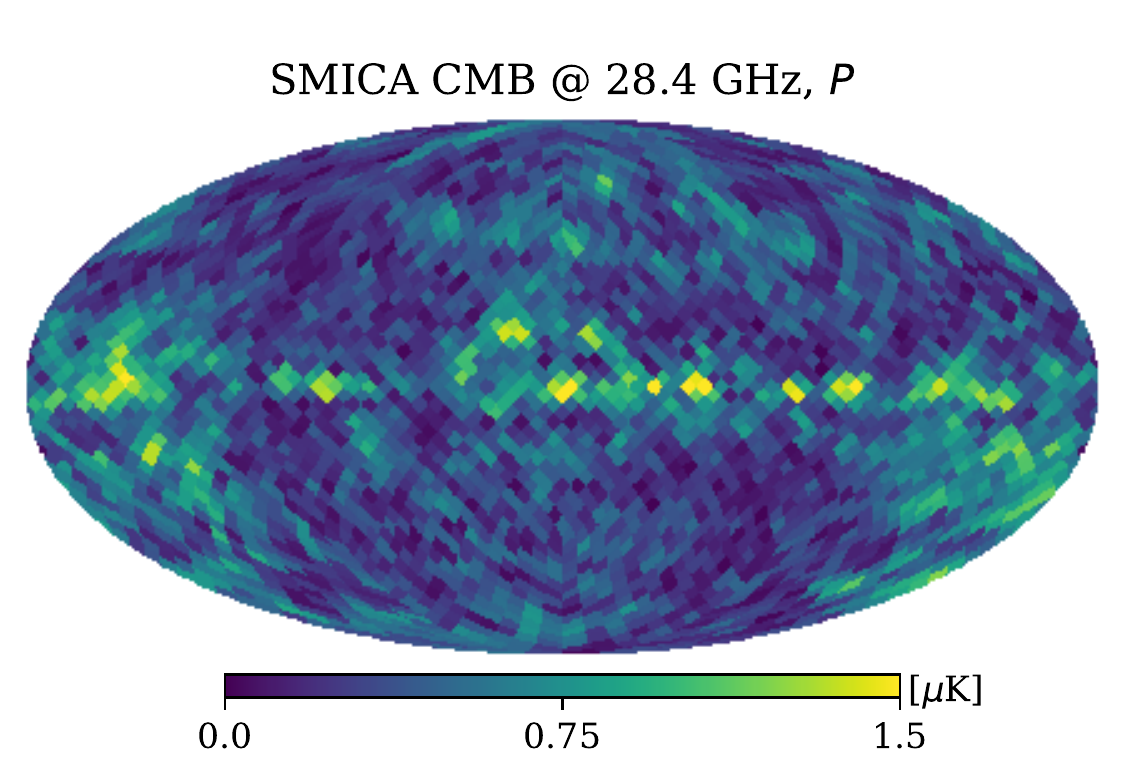}}
    \caption{The \textit{Planck} polarized intensity maps on a logarithmic colour scale (\textit{top row}), the uncertainty on the polarized intensities on a linear colour scale (\textit{middle row}), the Haslam total intensity map on a logarithmic colour scale(\textit{bottom left}), and the SMICA polarized intensity map in Rayleigh-Jeans brightness temperature at 28.4\,GHz on a linear colour scale (\textit{bottom right}). All the maps have been smoothed to $3.7^\circ$ FWHM and downgraded to $N_\textrm{side}=16$ resolution. The CMB fluctuations have been subtracted from the \textit{Planck} maps. The maps are in units of antenna temperature.}
    \label{fig:data_maps}
\end{figure*}

Joint analysis of multiple datasets is easiest if all maps are at a common resolution.
Therefore, after adding small noise realizations and subtracting the SMICA estimate of the CMB from the $Q$ and $U$ maps,
we deconvolve the beams from the maps and smooth to a common resolution of $3.7^\circ$ by reweighting the maps in $a_{\ell m}$ space.
$3.7^\circ$ is approximately the width of an $N_\mathrm{side}=16$ \textsc{HEALPix} pixel, which we downgrade the resolution of the maps to.\footnote{The pixel resolution of a \textsc{HEALPix} map is given by the $N_\textrm{side}$ parameter, the sky is broken into $12\times N_\textrm{side}^2$ equal-area pixels.}
The smoothed-and-downgraded Haslam, SMICA, and CMB-subtracted, smoothed-and-downgraded \textit{Planck} maps with their uncertainties are shown in Figure~\ref{fig:data_maps}.

We approximate the Haslam beam as a Gaussian with $\theta_\mathrm{FWHM}=56'$.
To smooth this map to $3.7^\circ$ resolution therefore requires a convolution with a Gaussian with $\theta_\mathrm{FWHM}=\sqrt{220'^2-56'^2}=212.8'$.
We approximate the beam in the SMICA map as a Gaussian with $\theta_\mathrm{FWHM}=10'$ and therefore smooth it to our target resolution of $220'$ by a convolution with a Gaussian with $\theta_\mathrm{FWHM}=219.8'$

For the \textit{Planck} surveys, 
we model the sky maps as being the true sky signal, convolved by an azimuthally symmetric beam and corrupted by additive white noise,
\begin{equation}
    y(\theta,\phi) = h\ast x (\theta,\phi) +n(\theta,\phi),
\end{equation}
where $y$ is the measured signal, $(\theta,\phi)$ are the angular coordinates, $h$ is the beam profile, $x$ is the true un-convolved sky signal and $n$ is Gaussian white noise.
We construct an estimator of $x$,
\begin{equation}
    \hat{x}(\theta,\phi) = (g\ast y)(\theta,\phi),
\end{equation}
where $g$ is an azimuthally symmetric function.

The Legendre transform of the function $g$ that provides the optimal least-squares-error estimator of $x$ is 
\begin{equation}
    G(\ell) = \frac{1}{H(\ell)}\frac{H(\ell)^2}{H(\ell)^2+N(\ell)^2/X(\ell)^2},
\end{equation}
where capital letters denote harmonic transforms.

In the limit of infinite signal-to-noise $G$ tends towards the simple inverse filter, $1/H$.
For $H$, we use the beam profiles in \citet{PlanckCollaboration2015e}.
Both $X$ and $N$ are not known a priori but must be estimated from the data themselves.
We approximate $X$ by fitting a power law to the angular power spectrum of the map between $\ell=10\--100$ and approximate $N^2$ as white noise from the high-$\ell$ part of the spectrum.

The deconvolution and smoothing operations are commutative and independent and so after deconvolution
we smoothed the maps to a Gaussian using the transfer function
\begin{equation}
    B(\ell) = \exp\left(-\frac{1}{2}\ell(\ell+1)\sigma^2\right),
\end{equation}
where $\sigma$ is the standard deviation of the Gaussian.

We downgrade the maps to lower $N_\mathrm{side}$ in $E-B$ space using the \textsc{HEALPix} \textsc{ud\_grade} subroutine before transforming back to $Q-U$.\footnote{The \textsc{ud\_grade} subroutine does not include parallel transport.}

After smoothing and downgrading, the variance maps need to be scaled by a correction factor. This factor is difficult to calculate analytically and so instead we estimate it using simulations.

From the smoothed and downgraded maps, in each pixel we construct $P(n)=\sqrt{Q(n)^2+U(n)^2}$ and approximate $\sigma_P(n) = \mathrm{max} (\sigma_Q(n),\sigma_U(n))$, where $n$ is the pixel number.

We also generate a simulated dataset on which to test the method.
We use the \textsc{PySM} s1 model \citep{Thorne2016} to simulate the Stokes $I$, $Q$ and $U$ diffuse Galactic synchrotron emission at 28.4\,GHz and $3.7^\circ$ resolution in an $N_\textrm{side}=16$ map.
Briefly this method uses the reprocessed, degree-scale smoothed Haslam $408\,\mathrm{MHz}$ map as an $I$ template \citep{Remazeilles2014}, and the 9-year \textit{WMAP} 23\,GHz maps smoothed to $3^{\circ}$ as $Q/U$ templates \citep{Bennett2013}.
Small scale features are added to these templates in \textsc{PySM} by generating Gaussian realisations from extrapolations of the angular power spectra to high $\ell$.
The synchrotron templates are extrapolated to other frequencies using a power-law frequency spectrum with a spatially varying spectral index taken from \citet{Miville-Deschenes2008} assumed to be the same in total intensity and polarization.

We then extrapolate the $Q$ and $U$ maps from 23\,GHz to 44.1\,GHz using a spatially constant temperature spectral index of $\beta=-3$. 
We add noise realisations to all the maps and create $P$ maps from the noisy maps of $Q$ and $U$.
The noise realisation for each pixel was generated by multiplying a number drawn from a standard normal distribution by $\sigma_P$ in that pixel from the \textit{Planck} noise maps.

We also need a simulated total intensity map to use as a prior.
For this we used the PySM total intensity synchrotron map at 28.4\,GHz, corrupted by a 5\,per cent multiplicative error in each pixel, again scaled to 44.2\,GHz with a temperature spectral index of -3, and apply a polarization fraction of 20\,per cent.
\section{Verification on simulated data} \label{sec:simsResults}
In this section we test our method on the simulated \textit{Planck} data.
We show in high signal-to-noise pixels the posterior estimates of the spectral index are approximately Gaussian and not significantly biased.
In lower signal pixels the method recovers posterior estimates that encode our lack of knowledge about the spectral index.
We describe how we use the mean shift clustering algorithm \citep{Comaniciu2002} implemented in \textsc{scikit-learn} \citep{Pedregosa2011} to group the pixels based on the posterior distributions of their spectral indices.
This allows us to identify where the spectral index has been well constrained.

\begin{figure*}
    \centering
    \subfloat[High signal-to-noise]{\includegraphics[width=\columnwidth]{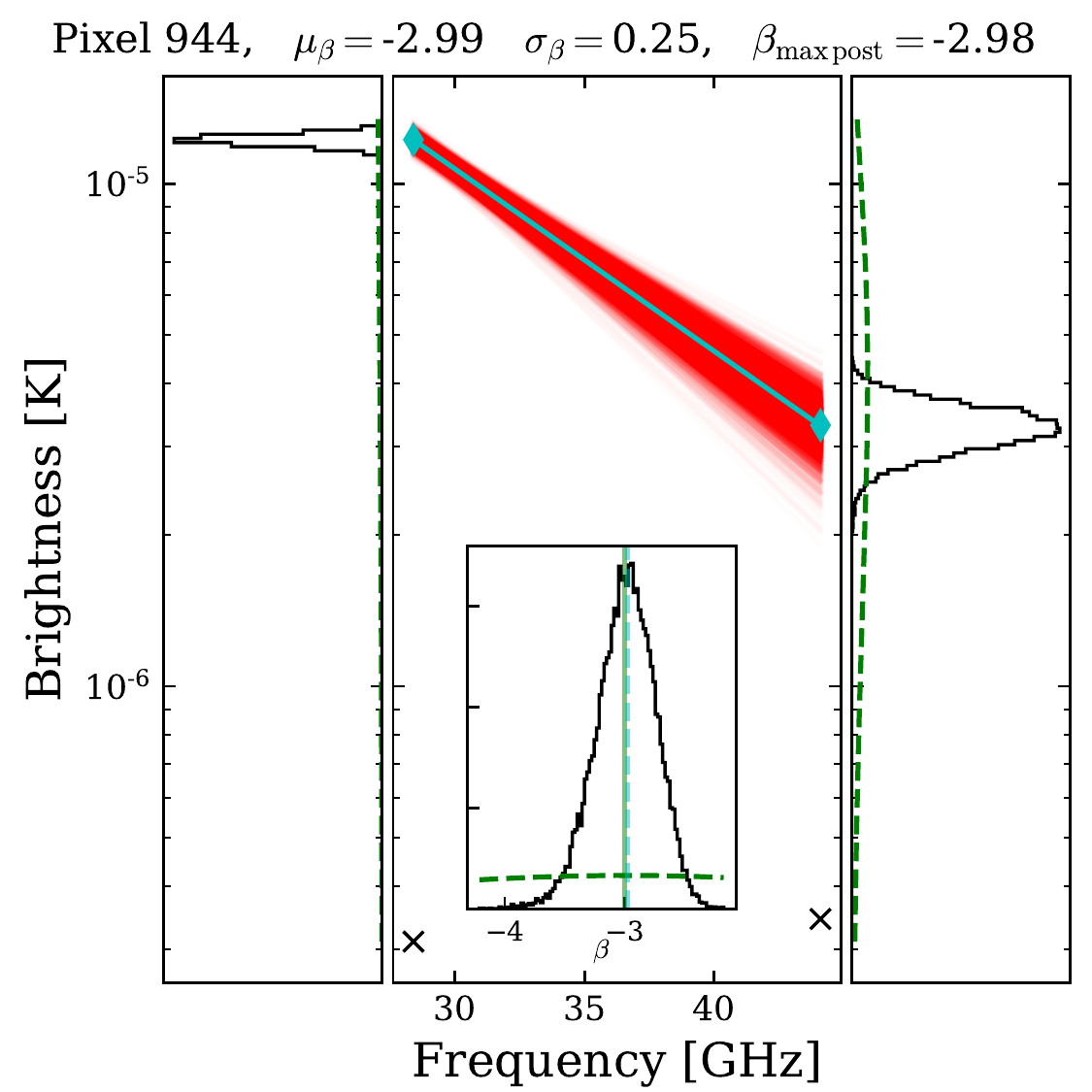}}
    \subfloat[Intermediate signal-to-noise]{\includegraphics[width=\columnwidth]{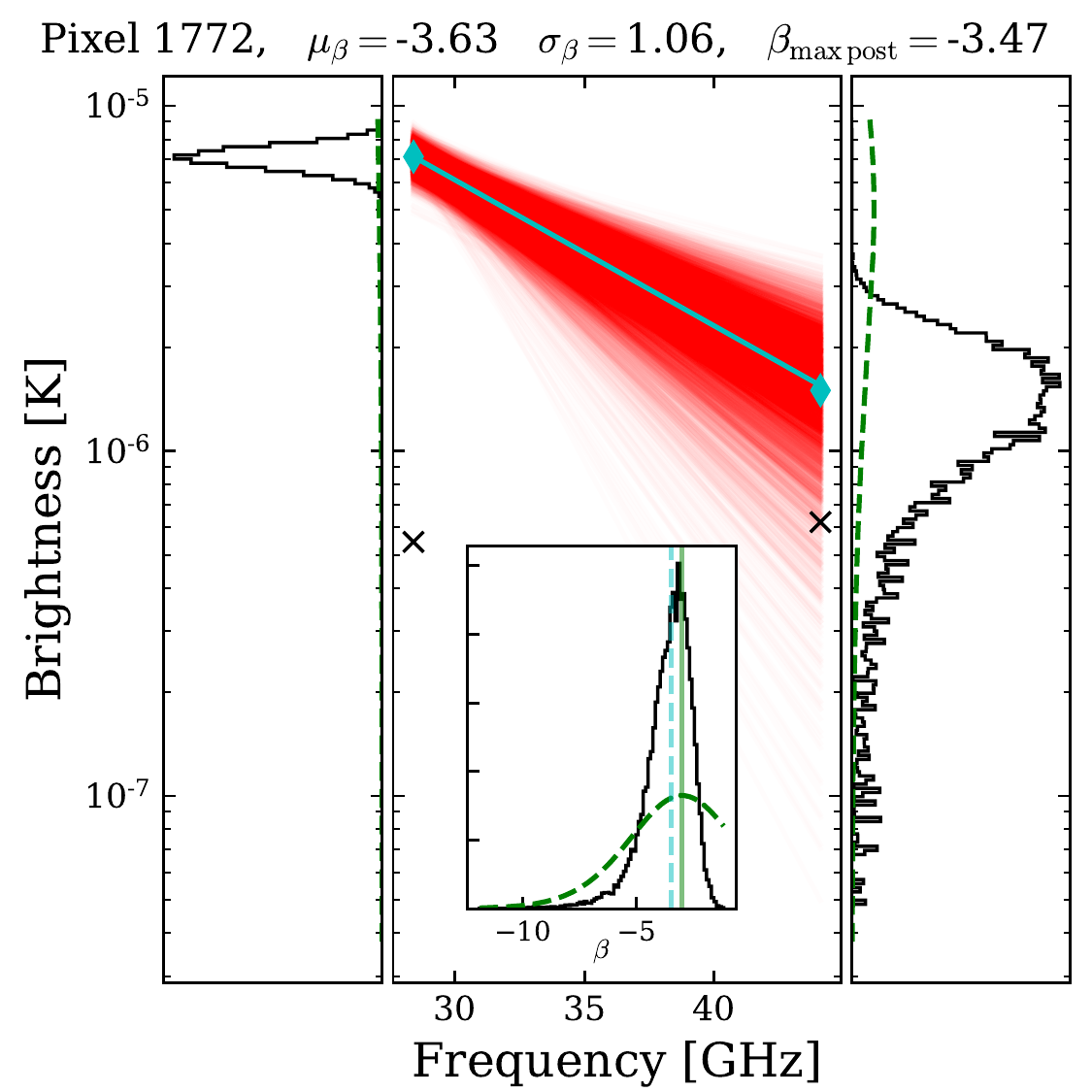}}
    \\
    \subfloat[Low signal-to-noise]{\includegraphics[width=\columnwidth]{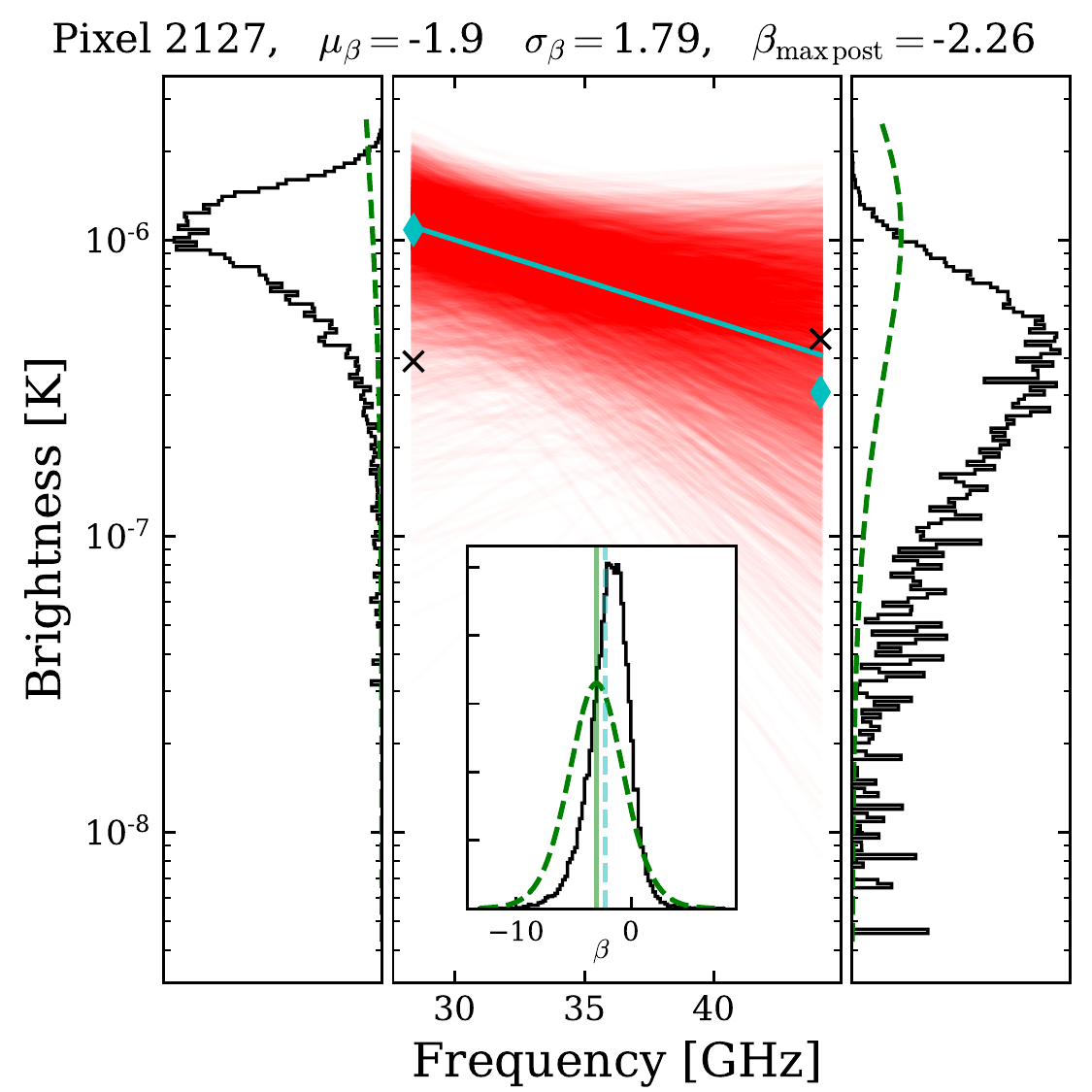}}
    \subfloat[Very low signal-to-noise]{\includegraphics[width=\columnwidth]{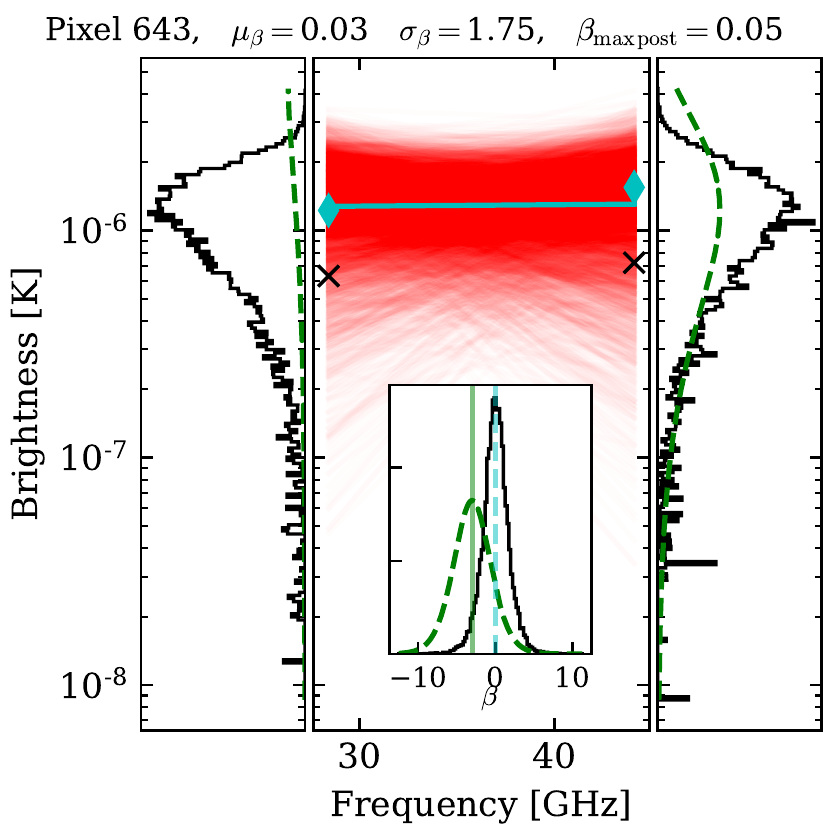}}
    \\
    \caption{Posterior distributions of the amplitude and spectral index parameters in four pixels from the simulated dataset representing signal-to-noise ratios that are high (\textit{top left}), intermediate (\textit{top right}), low (\textit{bottom left}) and very-low (\textit{bottom right}).
    The centre panels of each figure show the polarized intensities in both maps (\textit{cyan diamonds}), the noise levels (\textit{black crosses}), a thinned sample of the posterior distribution is plotted (\textit{red lines}) and the maximum posterior estimate is also shown (\textit{cyan line}).
    Marginalized posteriors of the two amplitude parameters are shown in the left and right side panels (\textit{solid black lines}) along with the priors (\textit{dashed green}).
    The inset axes in the central panels show the marginalized posterior distributions of the spectral index (\textit{solid black line}), the priors (\textit{dashed green line}), the maximum posterior estimate (\textit{vertical dashed cyan line}) and the true value of -3.0 (\textit{vertical solid green line}).
    }
    \label{fig:sim_pixelPost}
\end{figure*}

The posterior distributions of the amplitude and spectral index parameters for four pixels that represent high, intermediate, low and very low signal-to-noise signal-to-noise ratios are shown in Figure~\ref{fig:sim_pixelPost}.
The Kullback-Liebler divergences between the posterior and prior distributions on the spectral indices ($D_\mathrm{KL}$) are 0.66, 0.31, 0.22 and 0.87 in decreasing order of signal-to-noise, this pattern of large Kullback-Liebler divergences in pixels with either strong detections or such low signal-to-noise that the data can not constrain the spectral index is observed across the entire sky.
Above each panel we label the mean ($\mu_\beta$) and standard deviation ($\sigma_\beta$) of the posterior distribution of the spectral index.

\begin{figure}
    \centering
    \subfloat[Histograms of mean estimates of the spectral index for; all pixels (\textit{black line}), pixels with an uncertainty smaller than 1.5 (\textit{purple line}) and pixels with uncertainties less than 0.5 (\textit{green line}). The true spectral index is -3.0 (\textit{vertical dashed red line}).\label{fig:sim_1dHist_beta}]{\includegraphics[width=\columnwidth]{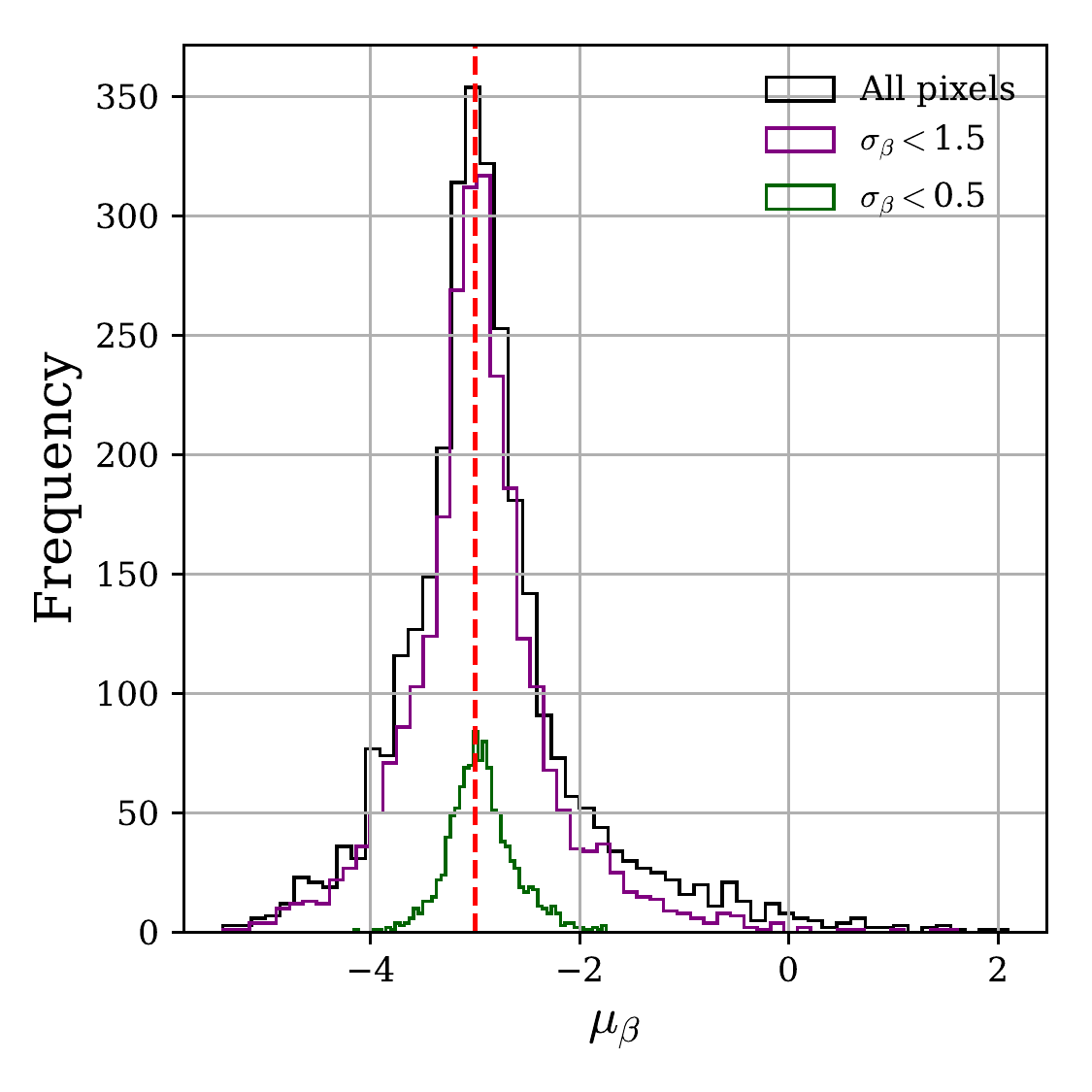}}\\
    \subfloat[Histogram of the normalized deviation of the spectral index estimates from the simulated data set (\textit{black line}) and the standard normal distribution (\textit{dashed red line}).\label{fig:sim1dHist_normVar}]{
    \includegraphics[width=\columnwidth]{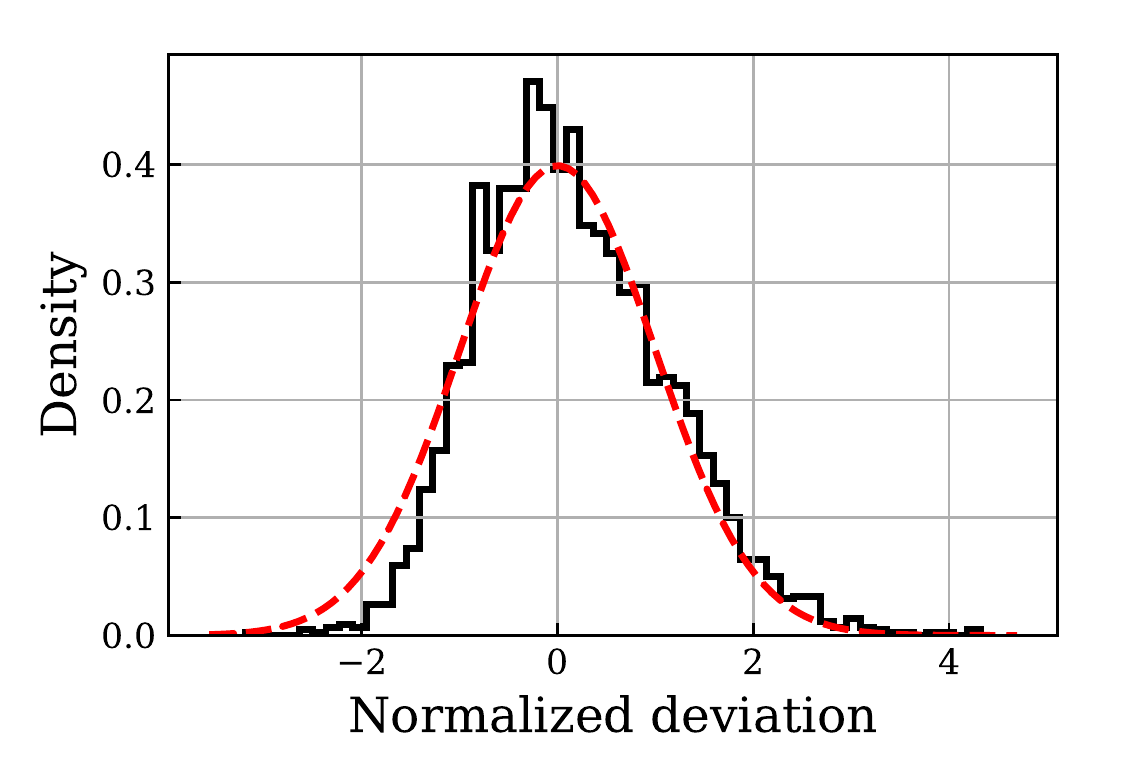}}
    \caption{Histograms of the mean posterior estimates of the spectral index from the simulated data (\textit{top}) and the normalized deviation of the estimates from the true value (\textit{bottom}).}
    \label{fig:sim_1dHist}
\end{figure}

Histograms of $\mu_\beta$, applying three cuts on $\sigma_\beta$,
are shown in Figure~\ref{fig:sim_1dHist_beta}.
All three histograms peak at the true synchrotron spectral index (\textit{dashed red line}).
The weighted average spectral index across the sky is
$\left<\beta\right>=-2.966\pm0.006$.
There is a slight excess of pixels with shallower spectral indices caused by pixels with very low signal-to-noise.
Given that neither the priors nor the likelihoods are Gaussians, there is no reason for the posteriors to be Gaussians.
However, we can still make that approximation and the normalized deviations of the estimates from the true value of -3.0 are shown in Figure~\ref{fig:sim1dHist_normVar}.
The distribution is close to the standard normal but slightly skewed towards positive deviations (shallower indices).

\begin{figure}
    \centering
    \includegraphics[width=\columnwidth]{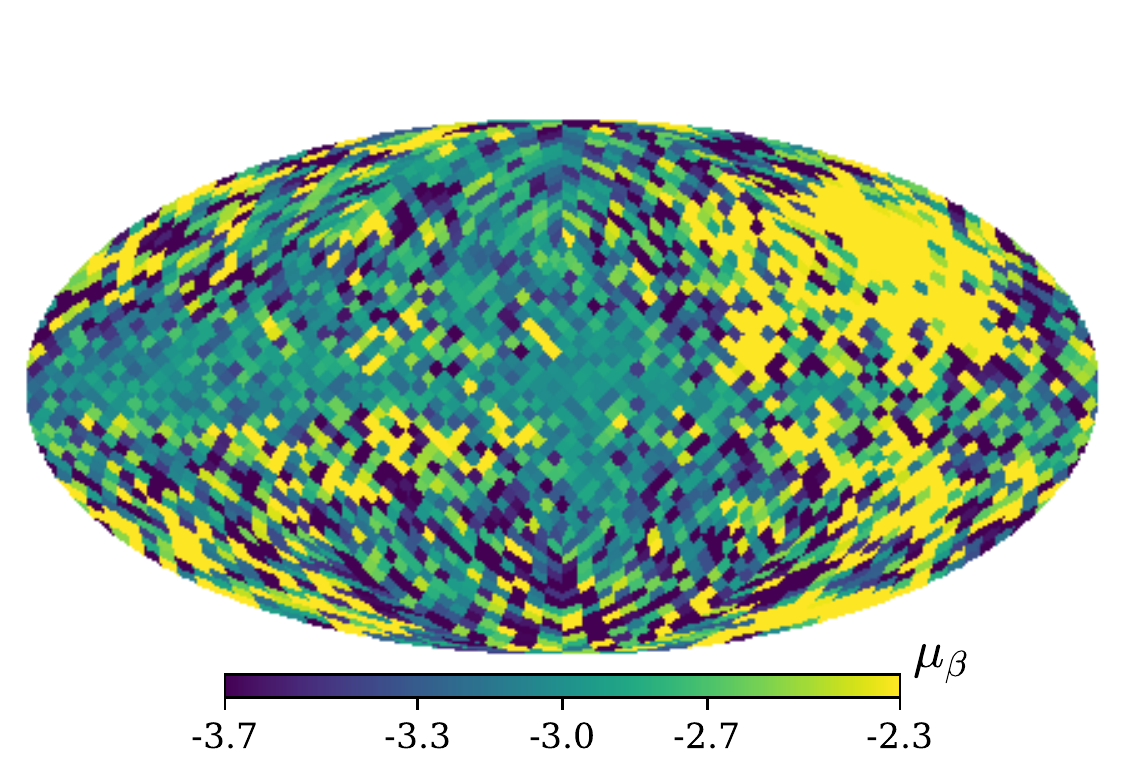}
    \includegraphics[width=\columnwidth]{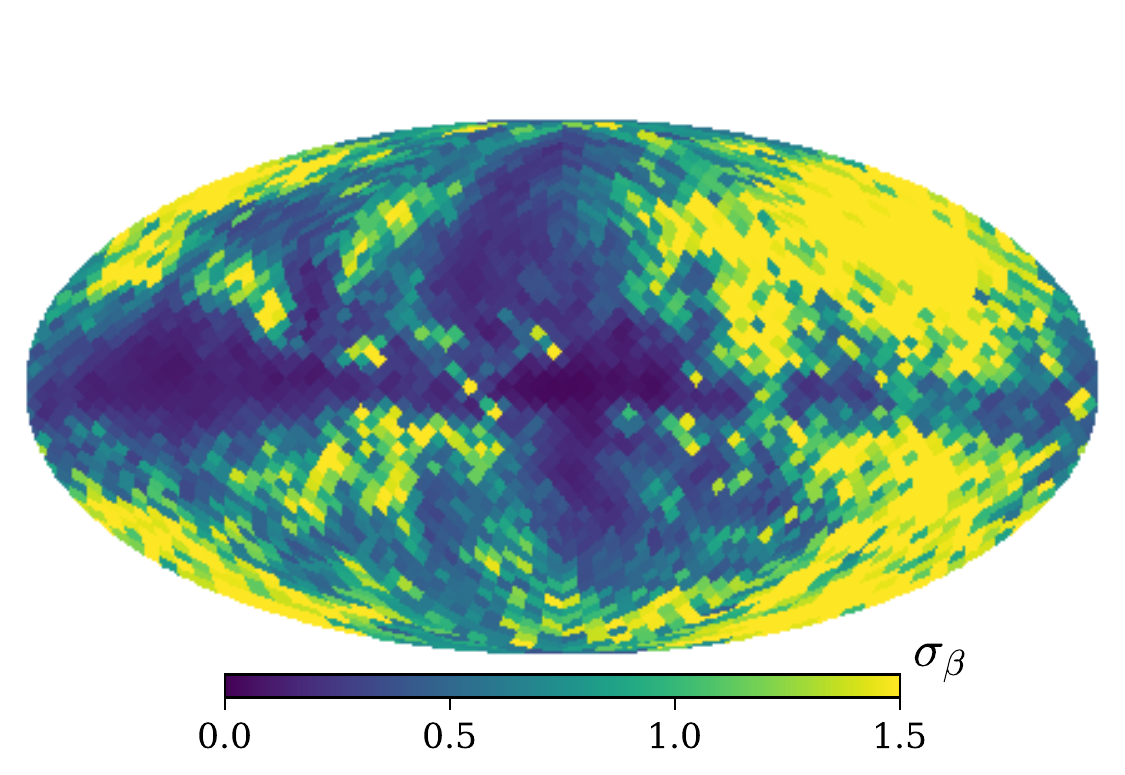}
    \includegraphics[width=\columnwidth]{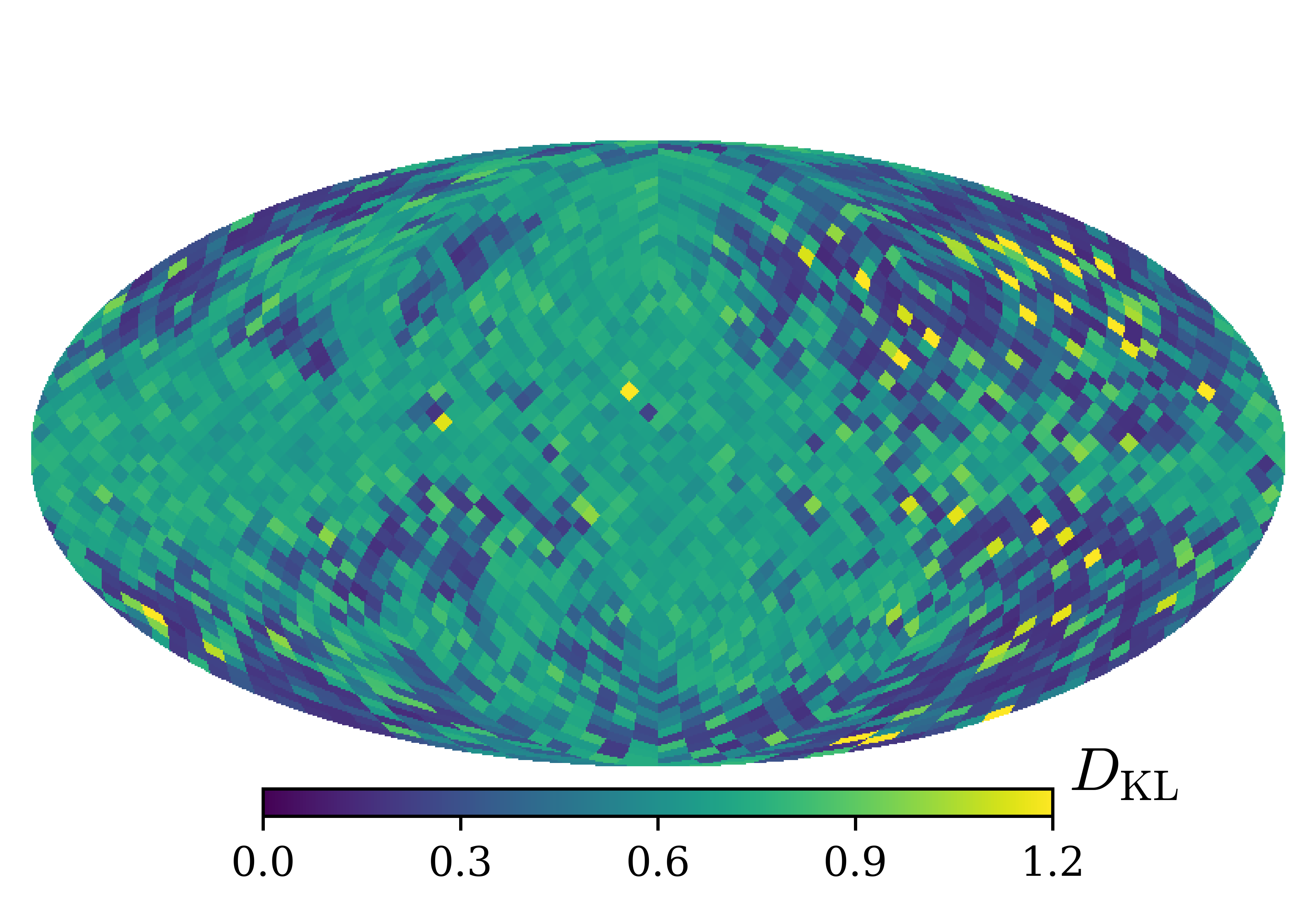}
    \caption{Mean (\textit{top}), standard deviation (\textit{middle}) and Kullback-Liebler divergence from the prior (\textit{bottom}) of the posterior estimates of the spectral index from the simulated data with true synchrotron spectral index of $\beta=-3.0$.
    }
    \label{fig:simResults_maps}
\end{figure}

The skewed distribution of the normalized deviation comes from the lowest signal-to-noise pixels.
Maps of the mean ($\mu_\beta$), standard deviation ($\sigma_\beta$) and Kullback-Liebler divergence from the prior ($D_\mathrm{KL}$), of the posterior distribution of the spectral index are shown in Figure~\ref{fig:simResults_maps}.
In low signal-to-noise regions, the posterior estimates of the spectral index have high uncertainties and in the faintest pixels (\textit{top right} region of the maps) the means of the posterior distributions are biased towards shallow spectral indices.

\begin{figure}
    \centering
    \includegraphics[width=\columnwidth]{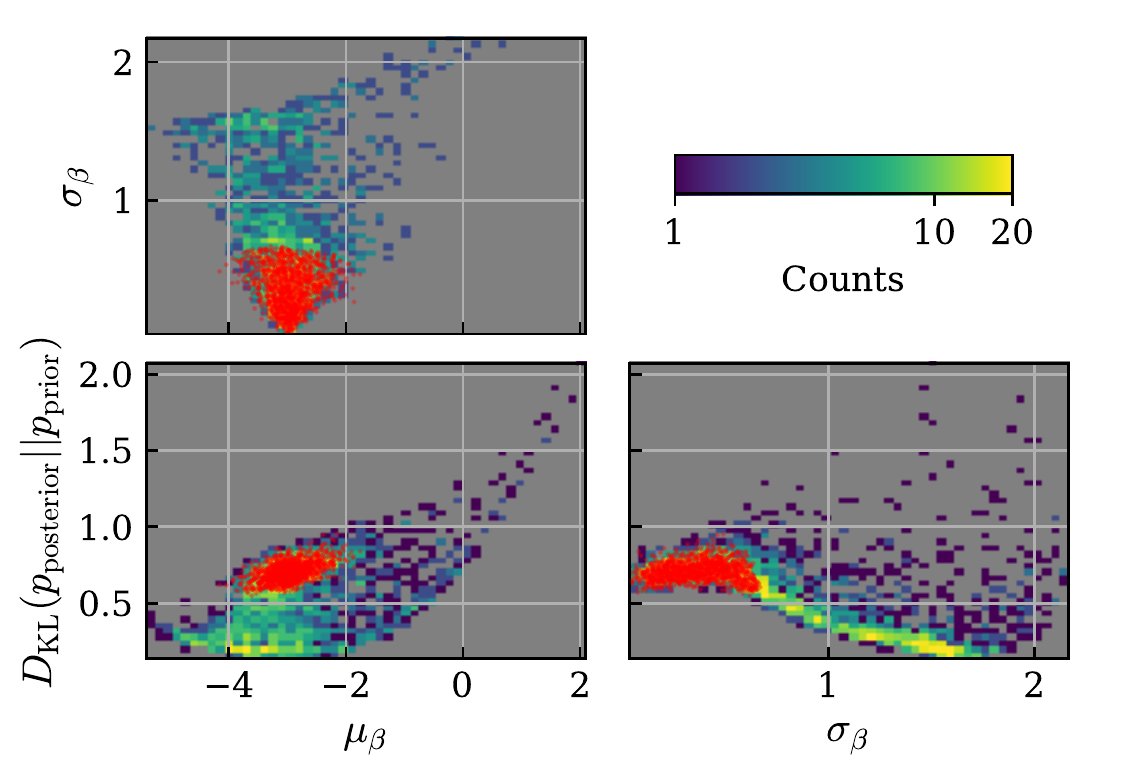}
    \caption{Results from the simulated data, 2-dimensional histograms of the estimated spectral indices, their uncertainties and the Kullback-Liebler Divergence between the prior and posterior spectral index distributions.
    The positions of pixels identified by the clustering algorithm are shown in \textit{red}, the details of which are described in Section~\ref{sec:simsResults}.}
    \label{fig:simResults_bsd_cluster}
\end{figure}

2-dimensional histograms of $\mu_\beta$, $\sigma_\beta$ and $D_\mathrm{KL}$ are shown in Figure~\ref{fig:simResults_bsd_cluster}.
There are at least two populations of pixels,
in the $D_\mathrm{KL}/\beta$ histogram there is an arc of pixels whose relative entropy increases with distance from the mean prior value and a second cluster with spectral index close to the true value and larger relative entropy ($D_\mathrm{KL}\simeq0.75$).
In the $D_\mathrm{KL}/\sigma_\beta$ histogram, most pixels lie along the line that is horizontal before a break at $\sigma_\beta\sim0.6$ where the relative entropy begins to decrease with increasing $\sigma_\beta$.

We use the mean shift clustering algorithm to cluster the pixels into groups based on where they are in $\mu_\beta$, $\sigma_\beta$ and $D_\mathrm{KL}$ space.
The mean shift algorithm starts walkers at the positions of the pixels and the walkers iteratively make steps towards regions of greater density \citep{Comaniciu2002}.
When the positions of the walkers have converged, those that end up close together are grouped.
The tunable parameter in this algorithm is the bandwidth that is used to calculate the local density.
Before clustering we scale all variables so that they span a range of 1. We then cluster using a bandwidth of 0.075.
The pixels belonging to the largest group (approximately half of the sky) have been identified in \textit{red} on the 2-dimensional histograms.
This cluster of pixels correspond to those where the spectral index has been detected and well constrained by the data.

Instead of simply calculating a weighted average spectral index,
by approximating the spectral index posterior distributions as Gaussians and assuming that the true spectral indices are themselves drawn from a Gaussian distribution,
we can fit for a mean value and the extra intrinsic scatter in the data that can not be accounted for by the uncertainties.
Across the entire sky the mean spectral index is $\left<\beta\right>=-2.963\pm0.006$ and the extra intrinsic scatter is $\sigma_\mathrm{intrinsic}=0.036\pm0.013$, i.e. consistent with zero.
In the pixels identified by the clustering algorithm we find $\left<\beta\right>=-2.970\pm0.006$ and $\sigma_\mathrm{intrinsic}=0.03\pm0.01$. 
These averages indicate that any bias on the mean of the posterior is small compared to the uncertainties.
The low intrinsic scatters, which are consistent with zero, show that the uncertainties are representative of the spread of the estimates.
\section{Results on real data} \label{sec:realResults}
\begin{figure*}
    \centering
    \includegraphics[width=\columnwidth]{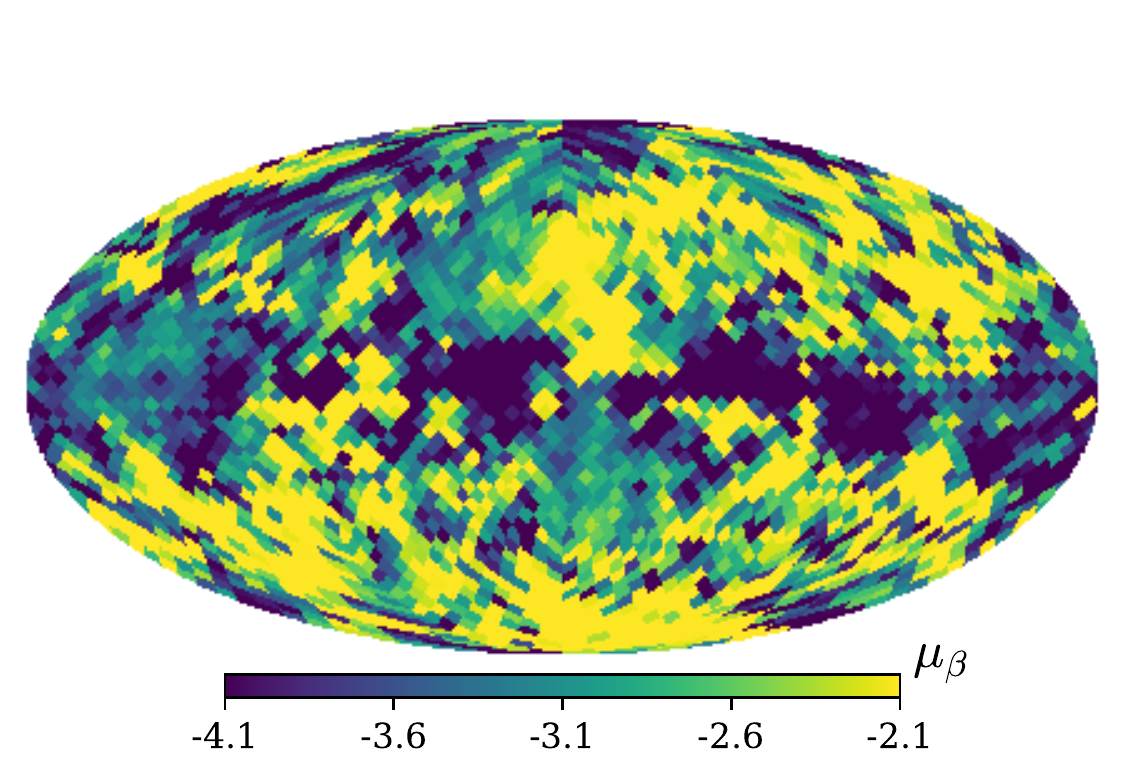}
    \includegraphics[width=\columnwidth]{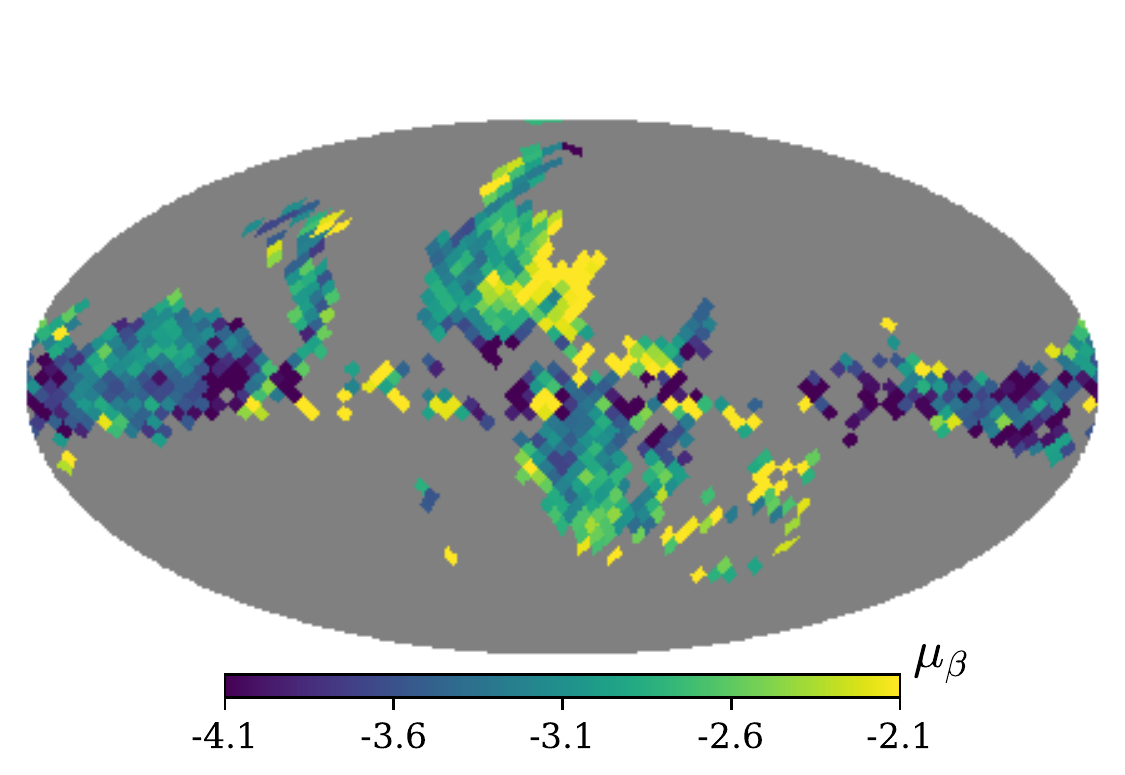}
    
    \includegraphics[width=\columnwidth]{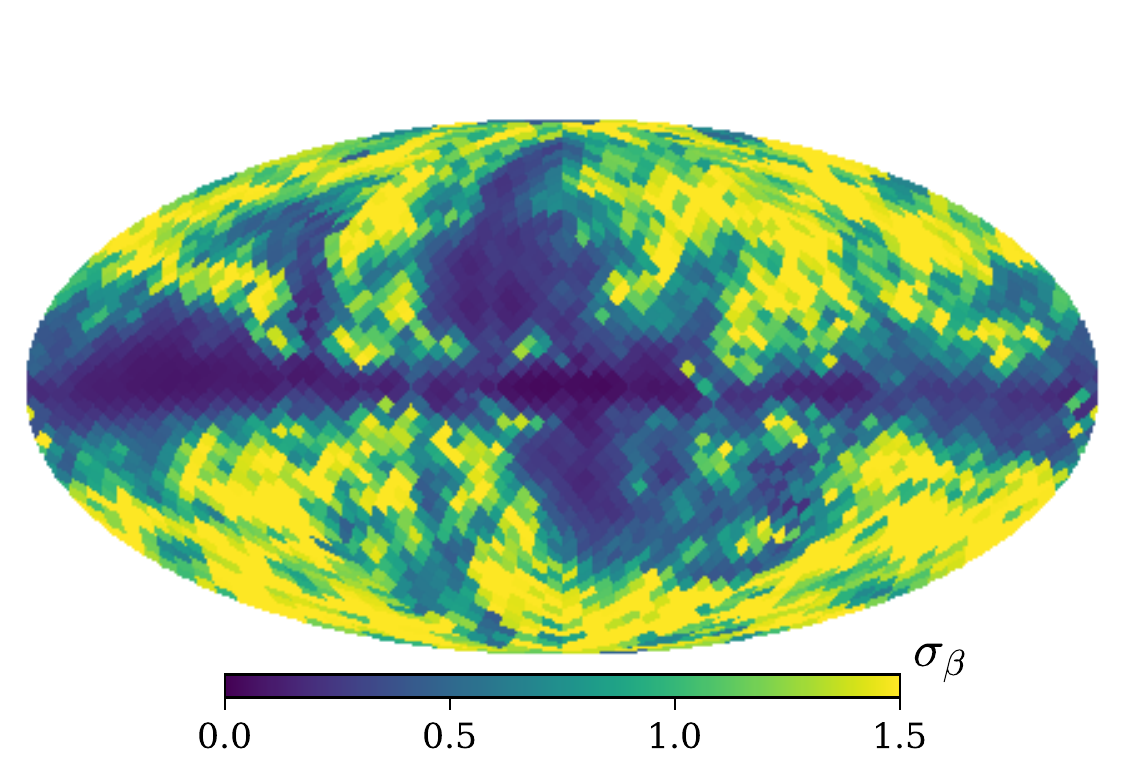}
    \includegraphics[width=\columnwidth]{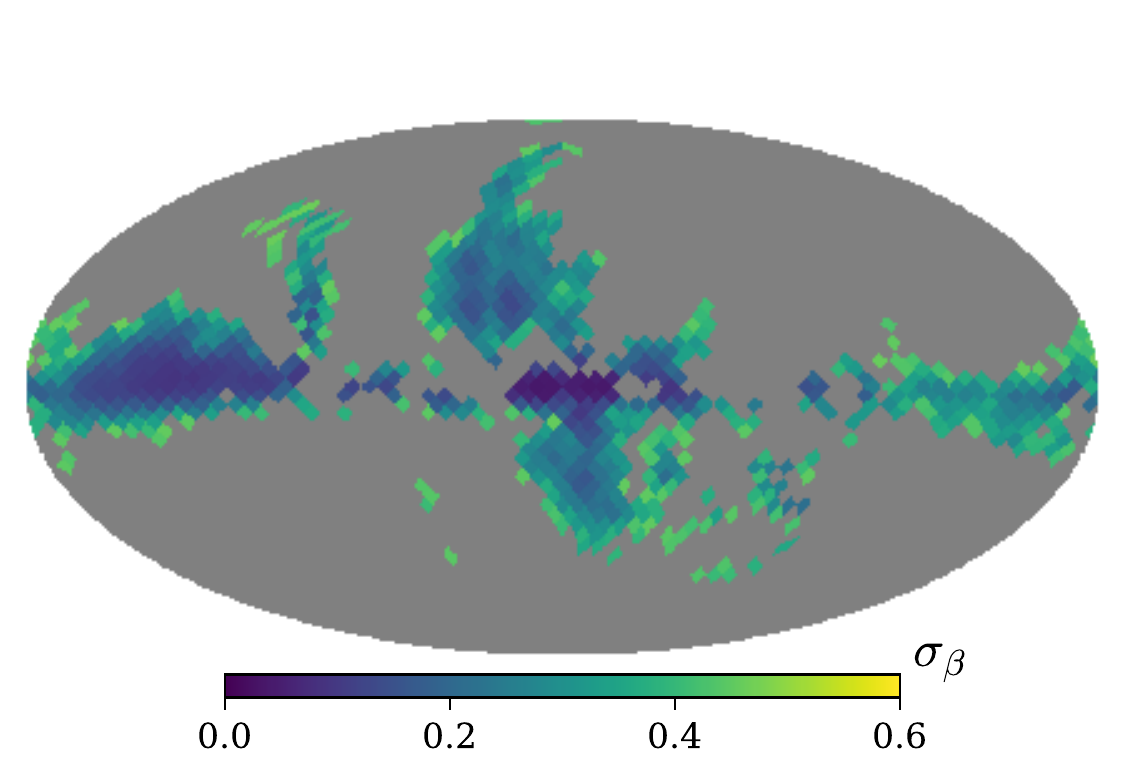}
    
    \includegraphics[width=\columnwidth]{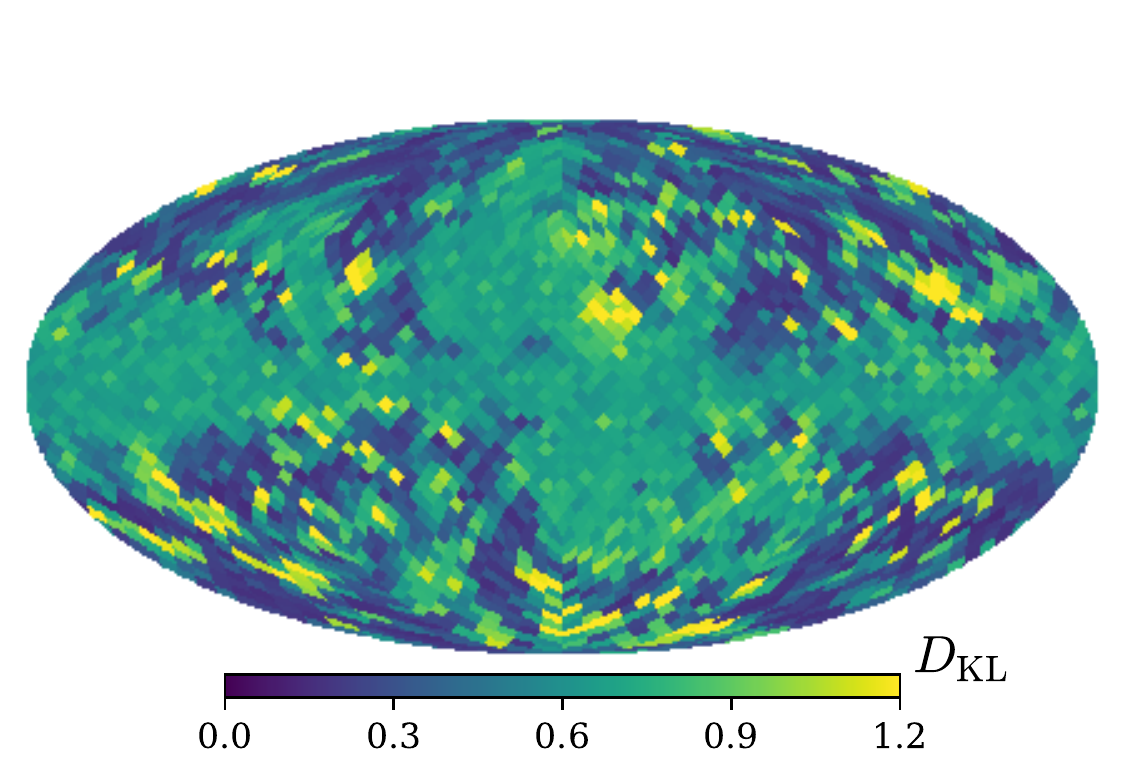}
    \includegraphics[width=\columnwidth]{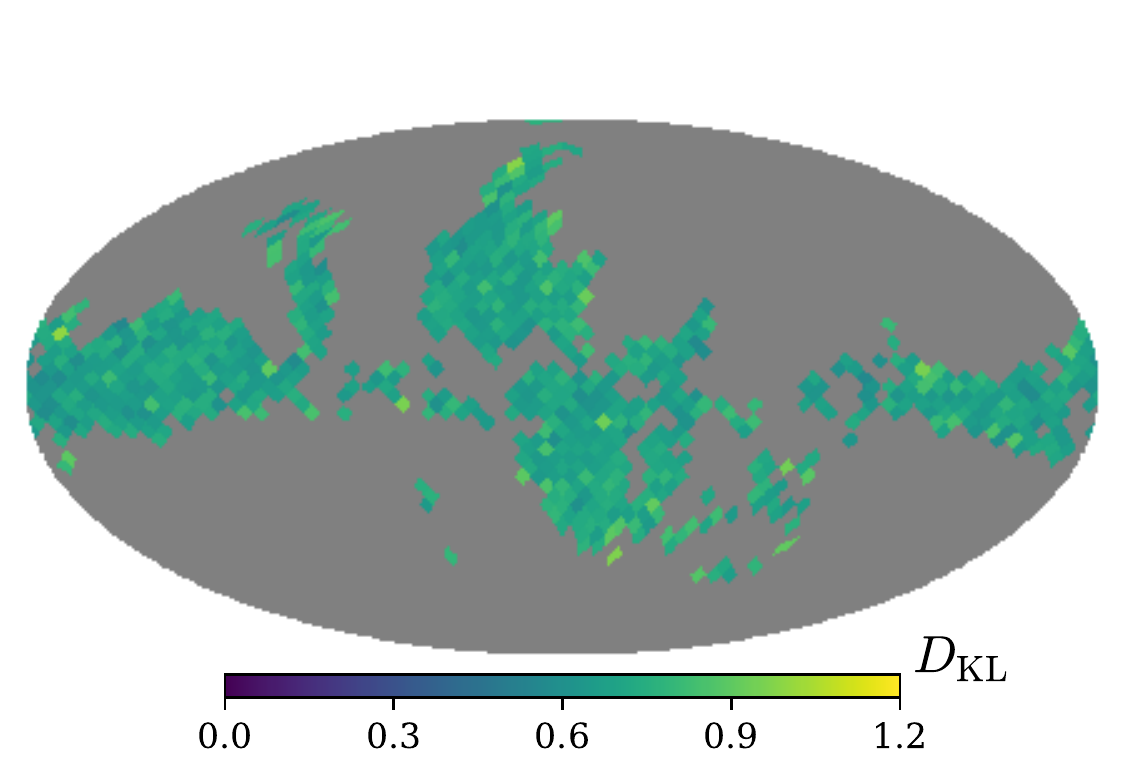}
    \caption{Mean (\textit{top}), standard deviation (\textit{middle}) and Kullback-Liebler divergence from the prior (\textit{bottom}) of the posterior estimates of the spectral index from the real data. On the \textit{right}, pixels not belonging to the group identified by the clustering algorithm have been masked.}
    \label{fig:realResults_maps}
\end{figure*}

Maps of the estimated spectral index between the real {\it Planck} 30 and 44\,GHz $P$ maps, the uncertainty of those estimates, and the Kullback-Liebler Divergence between the posteriors and the priors are shown in Figure~\ref{fig:realResults_maps}.

\begin{figure}
    \centering
    \includegraphics[width=\columnwidth]{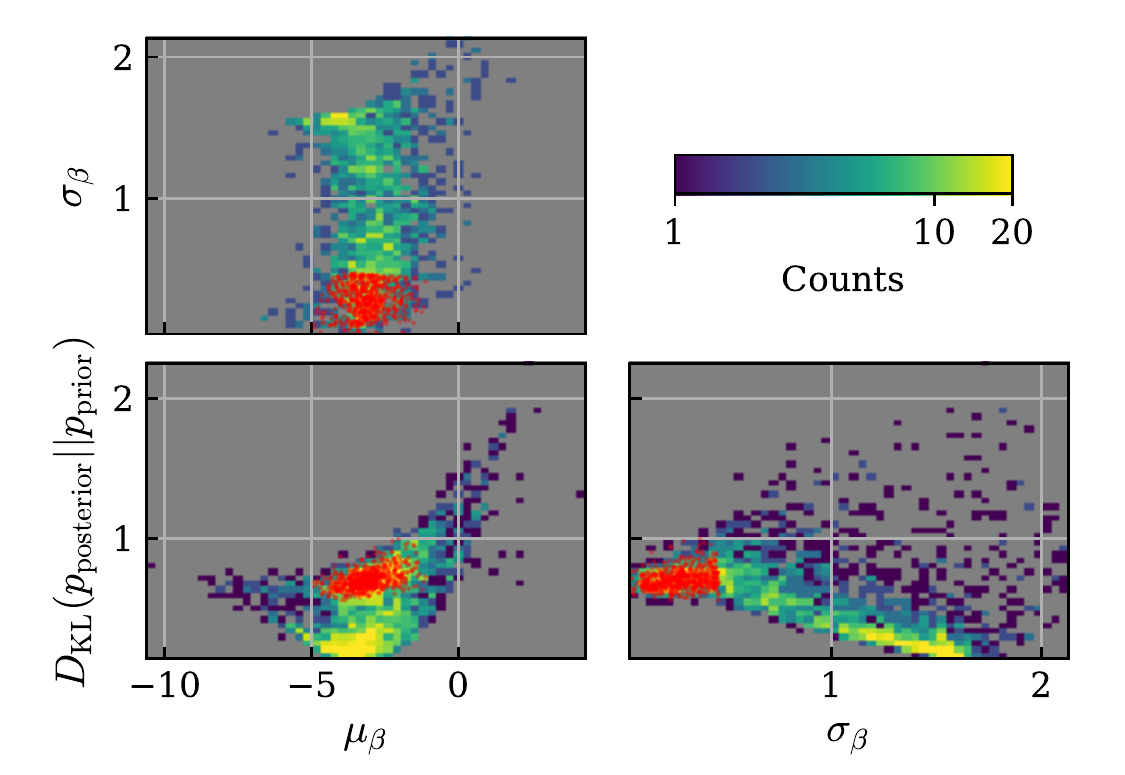}
    \caption{Results from real data, 2-dimensional histograms of the estimated spectral indices, their uncertainties and the Kullback-Liebler Divergence between the prior and posterior spectral index distributions.
    The positions of pixels identified by the clustering algorithm are shown in \textit{red}.}
    \label{fig:realResults_bsd_cluster}
\end{figure}

Once again, we use the mean shift clustering algorithm to group the pixels into clusters.
The largest cluster, which corresponds to those with good detections of the spectral index, contains 747 pixels (around one quarter of the sky).
2-dimensional histograms of $\mu_\beta$, $\sigma_\beta$ and $D_\textrm{KL}$ are shown in Figure~\ref{fig:realResults_bsd_cluster}, the locations of the pixels identified by the clustering algorithm  are shown in \textit{red}.
The histograms of real data show similar structures as the simulated data (cf. Figure~\ref{fig:simResults_bsd_cluster}).

\subsection{Average spectral index and intrinsic scatter}
The weighted average spectral index across the whole sky is -3.38.
This average is dominated by pixels along the Galactic plane with typically steep spectra and low uncertainty.
Along the Galactic plane we expect complicated frequency spectra due to the superposition of radiation from multiple synchrotron sources along the line of sight experiencing different levels of Faraday rotation as well as turbulent magnetic fields introducing frequency dependent fluctuations \citep[for example][]{Lazarian2016}.
The weighted average just including pixels that belong to the cluster is -3.09, and if then also ignoring pixels within $10^\circ$ of the Galactic plane the average is -2.96.

Again, by approximating the errors as Gaussian, we fit for both a mean value and the extra intrinsic scatter in the data that can not be accounted for by the errors alone.
The means and intrinsic scatters, along with values found by others, for large sky areas are listed in Table~\ref{tab:real_polBetas}.
Direct comparison between the averages found in this work and by others is difficult due to the different sky areas considered.
Estimates from data at lower frequencies typically find steeper spectral indices, perhaps indicating a shallowing of the spectrum with frequency.

Across the whole sky, we find $\left<\beta\right>=-2.99\pm0.03$ and $\sigma_\mathrm{intrinsic}=1.12\pm0.02$.
The mean and intrinsic scatter of just the pixels identified by the clustering algorithm are $\left<\beta\right>=-3.12\pm0.03$ and  $\sigma_\mathrm{intrinsic}=0.64\pm0.02$.
If we also neglect pixels within $10^\circ$ of the Galactic plane we find  $\left<\beta\right>=-2.92\pm0.03$ and  $\sigma_\mathrm{intrinsic}=0.48\pm0.02$.

These intrinsic scatters are significantly larger than the value found from simulated data, which had no intrinsic scatter.
When interpreting the intrinsic scatters, it is important to remember that we have approximated the extra spread of spectral indices as Gaussian.
This is likely not the case and the intrinsic scatter values are therefore just indicative of the variation, $\sim20$\,per cent from the mean.

\begin{figure}
    \centering
    \includegraphics[width=\columnwidth]{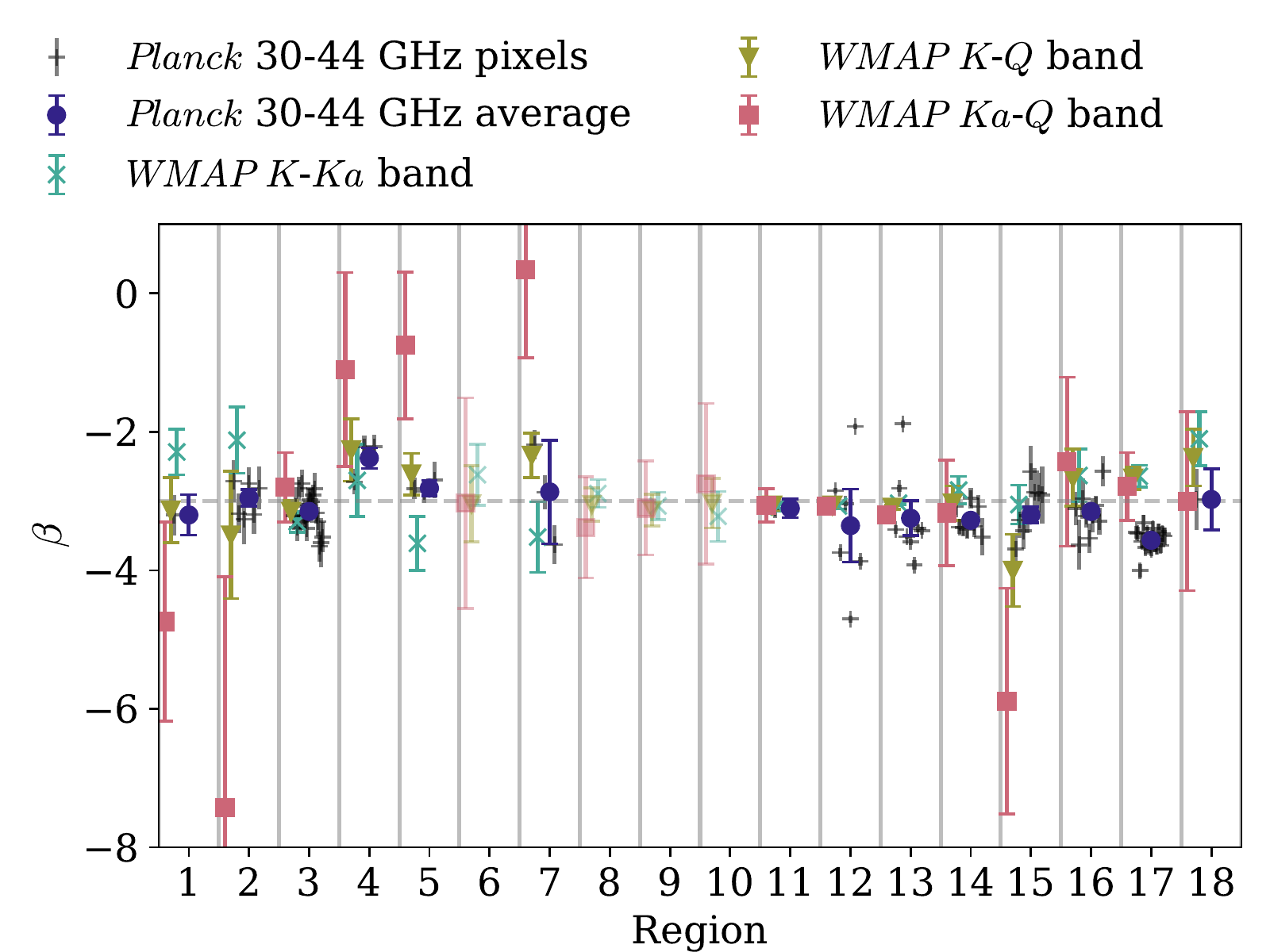}
    \caption{
    The polarized spectral indices that we find between the Planck 30 and 44 GHz surveys in the eighteen regions defined by \citet{Vidal2014}, along with the spectral indices found by \citet{Vidal2014} between the \textit{WMAP} low-frequency polarization surveys.
    The spectral index estimates of individual $N_\mathrm{side}=16$ pixels found in this work along with their 1-$\sigma$ error bars are shown by \textit{grey plusses}, the average values for each region are shown by the \textit{blue circles}.
    The \textit{cyan crosses} are the spectral indices between $K$- and $Ka$-bands.
    The \textit{yellow triangles} are the spectral indices between $K$- and $Q$-bands.
    The \textit{pink squares} are the indices between $Ka$- and $Q$-band.
    The horizontal \textit{dashed grey line} is at a spectral index of $\beta=-3$.
    The points have been greyed out in the regions that either do not enclose the central positions of any $N_\mathrm{side}=16$ pixels or do not contain any pixels identified by our clustering algorithm.}
    \label{fig:vidalCompare}
\end{figure}

\citet{Vidal2014} used $T-T$ plots to estimate the polarized spectral index between the \textit{WMAP} $K$-, $Ka$- and $Q$-band maps in eighteen regions.
For each region in our spectral index map, we fitted for an average value and extra intrinsic scatter.
The indices that we found along with those found by \citet{Vidal2014} are plotted in Figure~\ref{fig:vidalCompare} and listed in Table~\ref{tab:vidalCompare}.
We could not determine the average spectral index in four regions because three of them do not enclose the central coordinates of any $N_\mathrm{side}=16$ pixels (regions 6, 9 and 10) and one (region 8) contained no pixels identified by our clustering algorithm.
We could not estimate the intrinsic scatter in three of the regions (regions 1, 11 and 18) as they only contained a single pixel identified by our clustering algorithm.
The spectral indices found by us between \textit{Planck} 30 and 44\,GHz are generally consistent with those found by \citet{Vidal2014} between the \textit{WMAP} maps.

\begin{table*}
    \centering
        \caption{The average polarized spectral indices and intrinsic scatters that we find between the \textit{Planck} 30 and 44\,GHz surveys in the eighteen regions defined by \citet{Vidal2014}, along with the spectral indices found by \citet{Vidal2014} between the \textit{WMAP} low-frequency polarization surveys.
        The regions are rectangles in Galactic longitude and latitude, defined by their centre coordinates ($l_0$,$b_0$) and their widths.}
    \label{tab:vidalCompare}
    \begin{threeparttable}
    \begin{tabular}{lrrrr|rrr|rrr }
        \hline\hline
          &$l_0$ & $b_0$ & $l_\mathrm{width}$ & $b_\mathrm{width}$ & \multicolumn{3}{c}{$\beta$ from \citet{Vidal2014}}& \multicolumn{2}{c}{This work}\\
         Region &\multicolumn{4}{c}{[deg]} & $K$--$Ka$ & $K$--$Q$ & $Ka$--$Q$    &$\left<\beta\right>$ & $\sigma_\mathrm{intrinsic}$ & Notes\\
         \hline
1 & 5.0 & 75.0 & 30.0 & 10.0 & $-2.29\pm0.33$ & $-3.13\pm0.47$ & $-4.74\pm1.44$ & $-3.20\pm0.29$ & -- &$^a$ \\
2 & 25.0 & 57.0 & 10.0 & 14.0 & $-2.12\pm0.48$ & $-3.49\pm0.92$ & $-7.43\pm3.34$ & $-2.95\pm0.12$ & $0.02\pm0.06$\\
3 & 35.0 & 26.5 & 10.0 & 27.0 & $-3.30\pm0.15$ & $-3.15\pm0.16$ & $-2.80\pm0.50$ & $-3.15\pm0.05$ & $0.04\pm0.07$\\
4 & 20.0 & 27.5 & 10.0 & 5.0 & $-2.70\pm0.52$ & $-2.26\pm0.45$ & $-1.10\pm1.40$ & $-2.38\pm0.15$ & $0.09\pm0.23$\\
5 & 20.0 & 18.5 & 10.0 & 7.0 & $-3.61\pm0.39$ & $-2.61\pm0.30$ & $-0.75\pm1.06$ & $-2.81\pm0.11$ & $0.02\pm0.08$\\
6 & -6.5 & 21.5 & 9.0 & 7.0 & $-2.62\pm0.44$ & $-3.04\pm0.55$ & $-3.03\pm1.52$ & -- & -- & $^b$\\
7 & 0.0 & 13.5 & 10.0 & 7.0 & $-3.52\pm0.51$ & $-2.34\pm0.32$ & $0.34\pm1.27$ & $-2.87\pm0.75$ & $1.00\pm0.97$\\
8 & 12.0 & 7.5 & 10.0 & 5.0 & $-2.89\pm0.20$ & $-3.05\pm0.24$ & $-3.38\pm0.73$ & -- & -- & $^b$\\
9 & -6.5 & 7.5 & 7.0 & 5.0 & $-3.07\pm0.20$ & $-3.13\pm0.23$ & $-3.10\pm0.68$ & -- & -- & $^b$\\
10 & -14.5 & 11.0 & 7.0 & 8.0 & $-3.22\pm0.36$ & $-3.03\pm0.36$ & $-2.75\pm1.16$ & -- & -- & $^b$\\
11 & 32.0 & 0.0 & 24.0 & 8.0 & $-3.08\pm0.07$ & $-3.06\pm0.07$ & $-3.06\pm0.24$ & $-3.1\pm0.13$ & -- & $^a$\\
12 & 10.0 & 0.0 & 10.0 & 6.0 & $-3.07\pm0.04$ & $-3.06\pm0.04$ & $-3.07\pm0.12$ & $-3.35\pm0.53$ & $1.15\pm0.54$\\
13 & 345.0 & 0.0 & 20.0 & 6.0 & $-3.03\pm0.02$ & $-3.09\pm0.03$ & $-3.20\pm0.09$ & $-3.25\pm0.25$ & $0.70\pm0.22$\\
14 & 354.0 & -11.0 & 12.0 & 8.0 & $-2.84\pm0.20$ & $-3.03\pm0.25$ & $-3.17\pm0.76$ & $-3.28\pm0.05$ & $0.03\pm0.06$\\
15 & -2.0 & -26.5 & 16.0 & 13.0 & $-3.05\pm0.28$ & $-4.00\pm0.52$ & $-5.89\pm1.63$ & $-3.20\pm0.11$ & $0.05\pm0.12$\\
16 & 147.5 & 11.5 & 15.0 & 7.0 & $-2.63\pm0.38$ & $-2.66\pm0.41$ & $-2.43\pm1.22$ & $-3.15\pm0.07$ & $0.04\pm0.08$\\
17 & 135.0 & 0.0 & 30.0 & 6.0 & $-2.64\pm0.16$ & $-2.67\pm0.16$ & $-2.79\pm0.49$ & $-3.57\pm0.04$ & $0.12\pm0.05$\\
18 & 138.5 & -19.0 & 9.0 & 12.0 & $-2.10\pm0.39$ & $-2.37\pm0.41$ & $-3.00\pm1.29$ & $-2.98\pm0.44$ & -- & $^a$\\
\hline
    \end{tabular}
    \begin{tablenotes}
    \item $^a$ Region only contains one pixel identified by our clustering algorithm and so the intrinsic scatter can not be estimated.
    \item $^b$ Region does not contain the centres of any pixels identified by our clustering algorithm.
    \end{tablenotes}
    \end{threeparttable}
\end{table*}

\subsection{Angular power spectrum}
\begin{figure}
    \centering
    \includegraphics[width=\columnwidth]{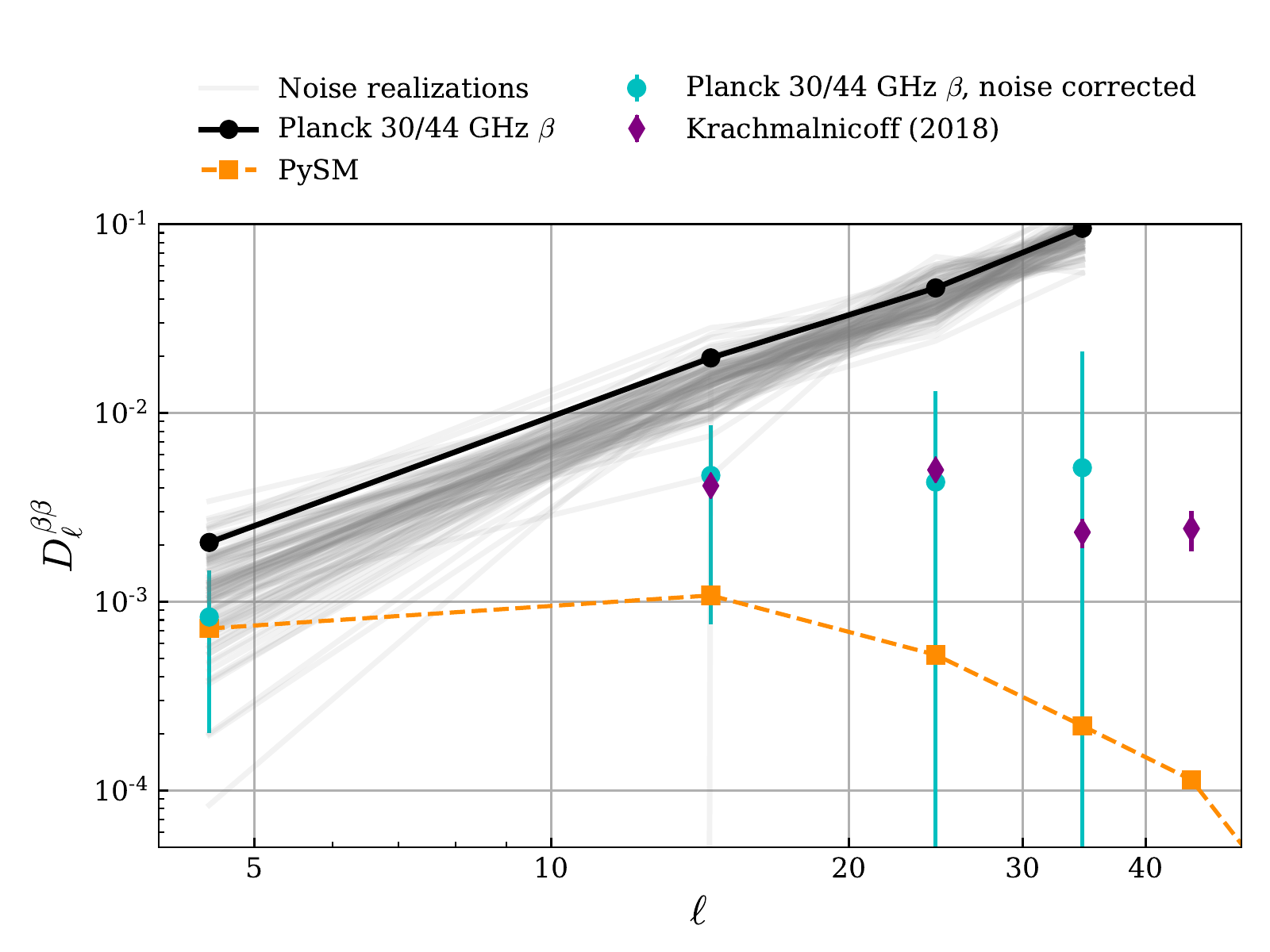}
    \caption{Angular power spectrum of the spectral index between the \textit{Planck} 30 and 44\,GHz polarized intensity surveys before (\textit{black}) and after (\textit{cyan}) correction for the noise.
    One hundred realizations of the noise are shown in \textit{grey}, the spectrum of the default spectral index map used by \textsc{PySM} is shown in \textit{dashed orange} and the spectral index found by the S-PASS collaboration is shown in \textit{purple diamonds}.}
    \label{fig:results_beta_spectrum}
\end{figure}

We use \textsc{PolSPICE}\footnote{\url{http://www2.iap.fr/users/hivon/software/PolSpice/}} \citep{Chon2003} to estimate the angular power spectrum of the polarized spectral index map, masking $25^\circ$ either side of the Galactic plane.
The spectrum is shown in Figure~\ref{fig:results_beta_spectrum} and
 is consistent with both the default spectral index map used by \textsc{PySM} \citep[from][]{Miville-Deschenes2008} and with the power spectrum of the spectral index between 1.4 and 28.4\,GHz found by \citet{Krachmalnicoff2018}.

In Figure~\ref{fig:results_beta_spectrum} the angular power spectrum of the mean spectral index map found in this work is shown in \textit{black}.
To estimate the impact of the noise on the spectrum we;
draw 100 samples of the spectral index from the converged MCMC chains for each pixel;
subtract the mean estimate of the spectral index in that pixel from the 100 samples;
form 100 maps from these mean-subtracted samples and estimate their power spectra.
These power spectra are shown in \textit{grey}.
At each multipole, the mean of these noise spectra is the noise bias and the standard deviation is an estimate of the uncertainty.
The noise de-biased spectrum and $1-\sigma$ error bars are plotted in\textit{cyan}.

For comparison we have also plotted the power spectrum found by \citet{Krachmalnicoff2018} and the power spectrum of the spectral index map used by \textsc{PySM} by default.
Both of these comparison spectra are consistent with the noise de-biased spectrum of the spectral indices found in this work.

\subsection{\textit{Fermi} bubbles/\textit{WMAP} haze}
\begin{figure}
    \centering
    \includegraphics[width=0.49\columnwidth]{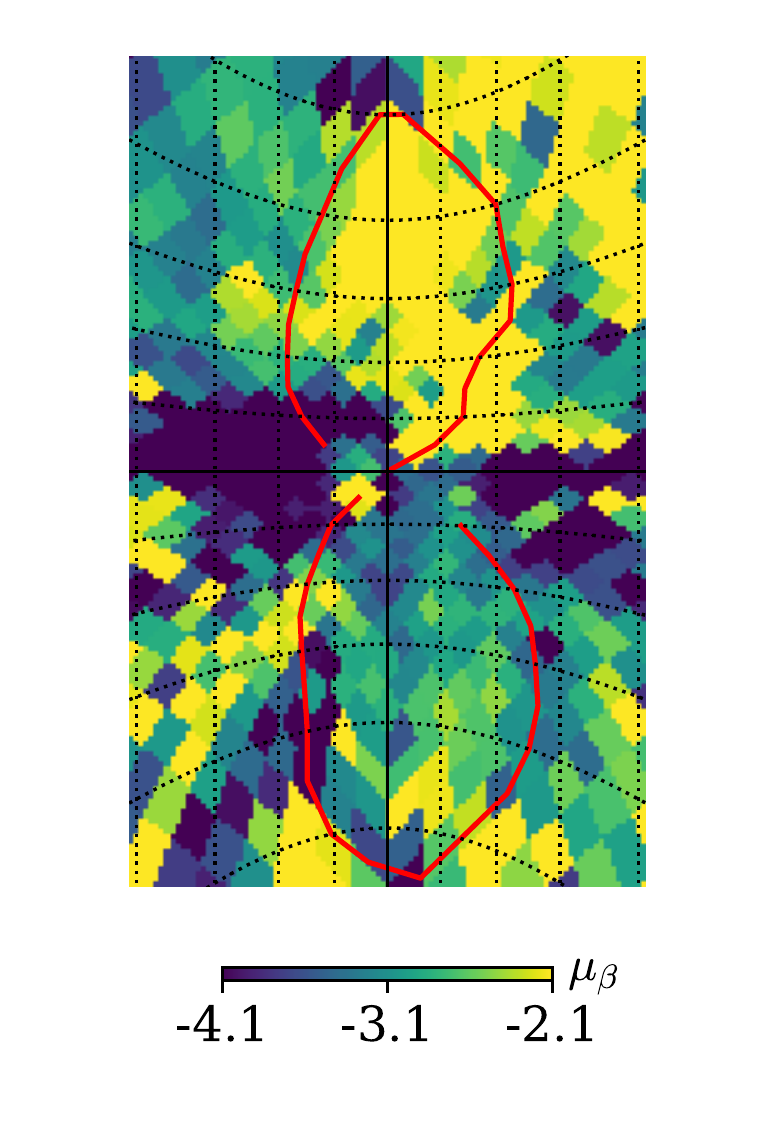}
    \includegraphics[width=0.49\columnwidth]{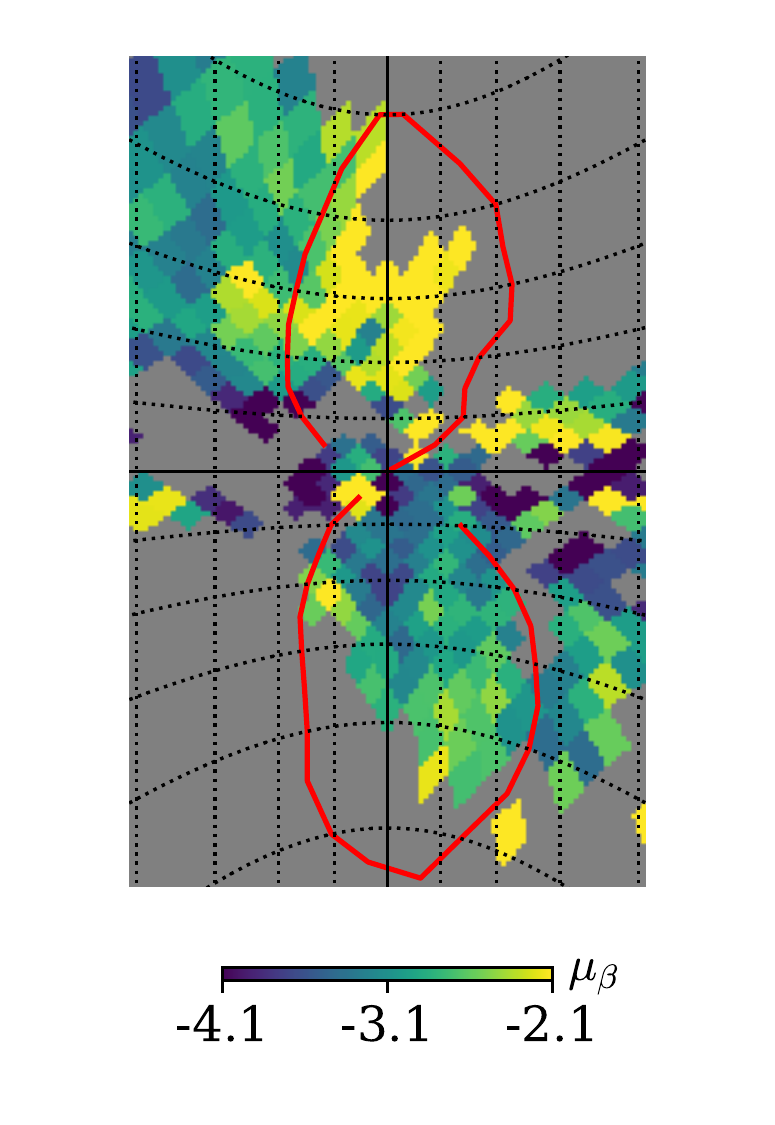}\\
    \includegraphics[width=0.49\columnwidth]{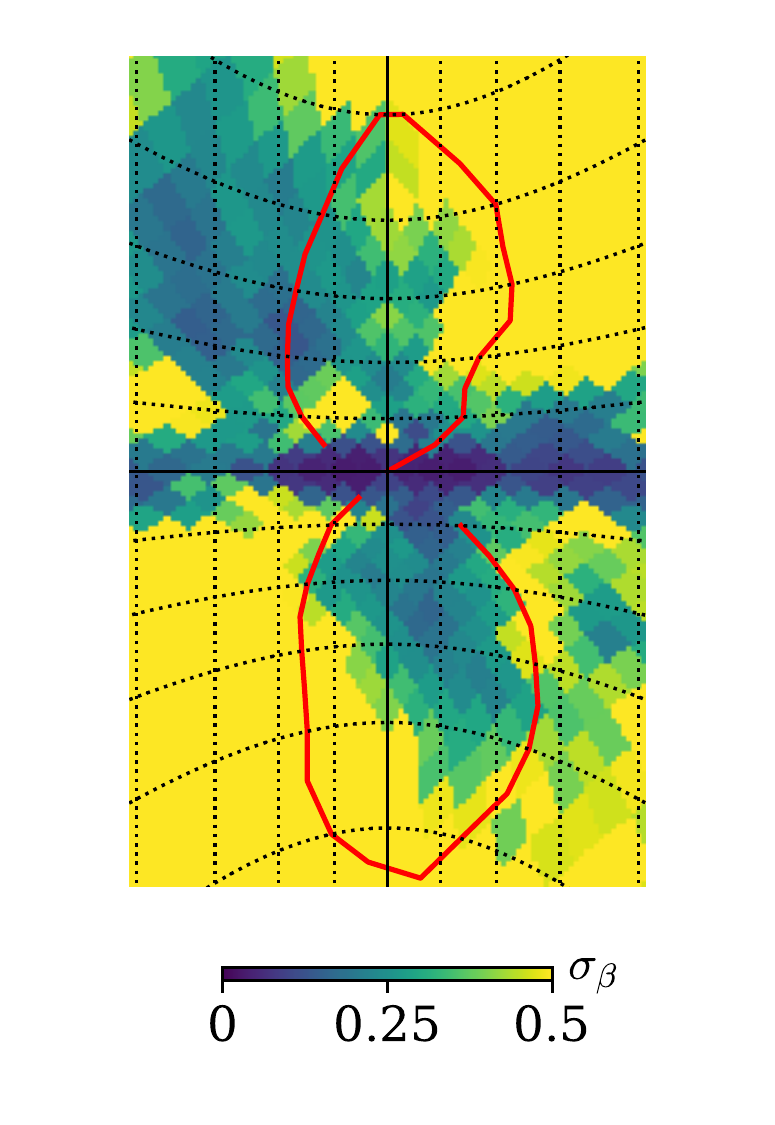}
    \includegraphics[width=0.49\columnwidth]{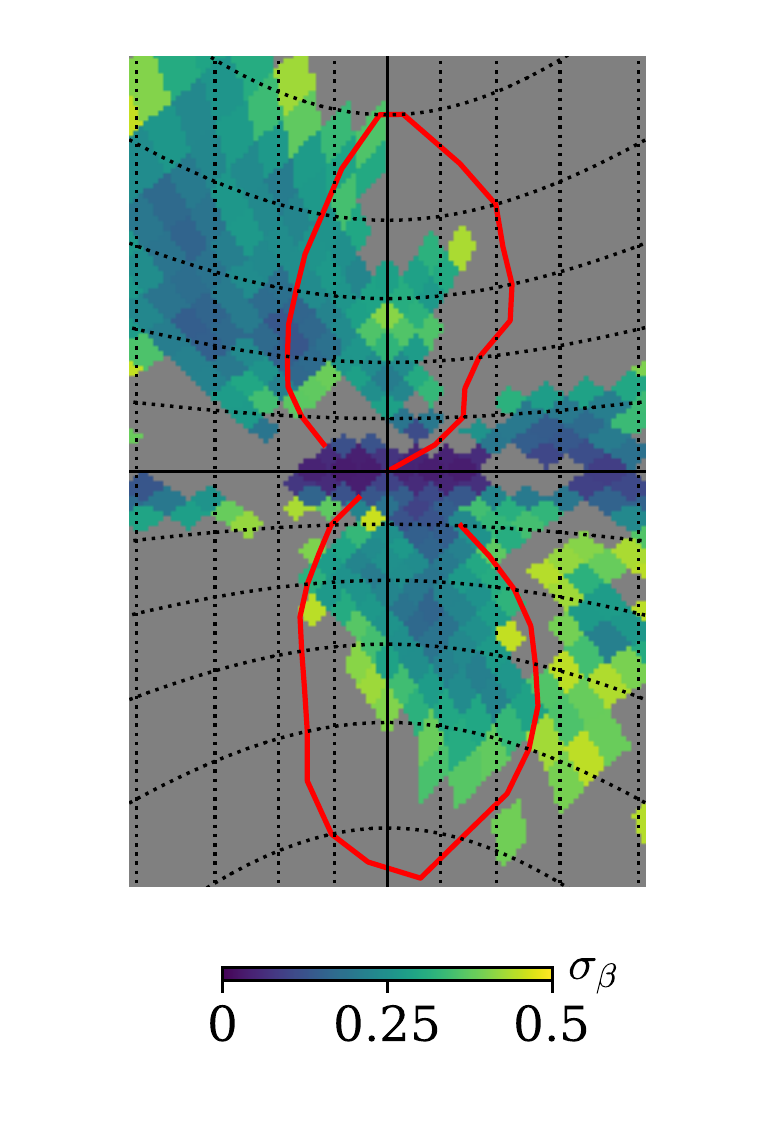}
    \caption{Gnomonic projections of the mean (\textit{left}) and standard deviation (\textit{right}) of the posterior estimates of the spectral index from real data.
    In the \textit{left column} no pixels have been masked, in the \textit{right column} pixels not identified by the clustering algorithm have been masked out.
    The projections are directed at the Galactic centre, the graticules show constant latitudes and longitudes separated by $10^\circ$.
    The \textit{red} lines mark the edges of the \textit{Fermi} bubbles as defined by \citet{Su2010}.
    }
    \label{fig:fermiBubbles}
\end{figure}

There is a region north of the Galactic centre with high-significance detections of shallow spectral indices.
A gnomonic projection of this region is shown in Figure~\ref{fig:fermiBubbles}.

Spatially the shallow spectrum emission correlates with the location of the northern \textit{Fermi} bubble (the coordinates defining the Fermi bubbles are from \citet{Su2010} and outlined in \textit{red}).
The average spectral indices are different in the north and south bubbles.
In the north bubble, only including pixels identified by the clustering algorithm, we find $\left<\beta\right>=-2.36\pm0.09$ with an intrinsic scatter of $\sigma_\mathrm{intrinsic}=0.63\pm0.07$.
In the south bubble, again only including pixels identified by the clustering algorithm, we find $\left<\beta\right>=-3.00\pm0.05$ and $\sigma_\mathrm{intrinsic}=0.35\pm0.04$.
Combining both regions together we find $\left<\beta\right>=-2.72\pm0.06$ and $\sigma_\mathrm{intrinsic}=0.57\pm0.04$.

For comparison, \citet{PlanckCollaboration2012a} included a diffuse 2-dimensional Gaussian centred on the Galactic centre as a template for this emission during template fitting.
They found a spectra index for the haze component in total intensity of $\beta=-2.56\pm0.05$, which is remarkably consistent with the average spectral index we find across both bubbles including all pixels (not just those identified by the clustering algorithm), $\left<\beta\right>=-2.57\pm0.08$ and $\sigma_\mathrm{intrinsic}=1.06\pm0.06$.

To determine whether the average spectral indices in the north and south bubbles are different we compute the Bayes factor.
The Bayes factor is the ratio of the evidence (or marginalized likelihood) for the case where the true averages are different to evidence for the case where the true averages are the same.
Bayes factors greater than unity are evidence in favour of the averages being different, factors less than unity are evidence for the averages being the same.
\citet{Jeffreys1983} suggests the following heuristic for interpreting Bayes factors;
Bayes factors between 1--3 are barely worth mentioning, Bayes factors between 3--10 are substantial, factors between 10--30 are strong evidence, factors between 30--100 indicate very strong evidence, and factors greater than 100 are decisive evidence.
The Bayes factor in this instance is $3\times10^7$.


The shallow spectral indices that we have found in the northern bubble could be caused by several effects and not all of them are astrophysical.
Inaccurate $\sigma_P$ values, incorrect CMB subtraction, $I$ to $P$ leakage at 44\,GHz,  depolarization at 30\,GHz, zero-level errors in the prior, anomalously highly-polarized free-free emission or shallow spectrum synchrotron emission could be responsible.
We discuss each of these in turn below.

\subsubsection{Inaccurate $\sigma_P$ values}
The $\sigma_P$ values for the \textit{Planck} maps could underestimate the true noise levels in the maps.
To determine the impact of using higher noise levels on the estimated spectral indices, we scaled the $\sigma_P$ values for both \textit{Planck} maps by a factor of two and a factor of ten, estimated the spectral index in each pixel and found the average in the \textit{Fermi} bubbles.
The values are listed in Table~\ref{tab:lobes_sigma_P_scaling}.

\begin{table*}
    \centering
    \caption{The mean spectral indices in the north and south bubbles and the evidence that they are different.
    In the first column, when not subtracting an estimate of the CMB from the maps and instead marginalizing over the CMB amplitude.
    In the second column, when subtracting an 8.9\,K monopole from the Haslam map that forms the weakly informative prior.
    In the remaining columns,
    after scaling the $\sigma_P$ noise levels of the \textit{Planck} maps by factors of 1.0, 2.0 and 10.0.
    The main results in this paper correspond to the third column, i.e. with a $\sigma_P$ scaling factor of unity and with the SMICA CMB estimate subtracted from the \textit{Planck} polarization maps and no Haslam monopole subtracted from the prior.
    }
    \label{tab:lobes_sigma_P_scaling}
    \begin{tabular}{r|rrrrr}
        \hline\hline
        & No CMB& Remove Haslam & Main result &\multicolumn{2}{c}{$\sigma_P$ scaling factor} \\
        & subtraction&monopole &from this work & 2.0 & 10.0 \\
        \hline\\
        $\left<\beta\right>^\mathrm{north\,bubble}$
        & $-2.47\pm0.10$ &$-2.37\pm0.09$& $-2.36\pm0.09$ & $-2.52\pm0.10$ & $-3.05\pm0.23$ \\
        $\left<\beta\right>^\mathrm{south\,bubble}$ 
        & $-3.09\pm0.05$ &$-3.03\pm0.05$& $-3.00\pm0.05$ & $-3.18\pm0.06$ & $-3.60\pm0.22$\\
        Bayes factor
        & $4\times10^5$ &$8\times10^7$& $3\times10^7$ & $3\times10^6$ & 4 \\
        $\left<\beta\right>^\mathrm{both\,bubbles}$ 
        & $-2.82\pm0.06$ &$-2.60\pm0.08$& $-2.72\pm0.06$ & $-2.88\pm0.06$ & $-3.33\pm0.16$ \\
        \hline
    \end{tabular}
\end{table*}

Without scaling the noise levels, the evidence of the difference is decisive with a Bayes factor of $3\times10^7$, scaling the noise levels by a factor of two only marginally decreases the evidence to $3\times10^6$, which is still decisive evidence for the averages being different.
The noise has to be scaled by very large factors to make the significance of the difference drop substantially.
When the noise is scaled by a factor of 10 the Bayes factor falls to around 4, just considered substantial by \citet{Jeffreys1983}.
We do not expect the noise levels on the maps to be wrong by such a large factor, but even if they are we still find substantial evidence for the average spectral index in each bubble being different.

Unsurprisingly, we find that as the noise scaling increases the combined average spectral index across both bubbles moves closer to the prior ($-3.11$).

\subsubsection{CMB subtraction}
The SMICA estimate of the CMB will be contaminated by residual foregrounds, particularly close to the Galactic plane.
We re-ran the fitting on \textit{Planck} maps without the CMB subtracted and instead, to account for the fluctuations in the CMB, 
marginalized over the CMB amplitude in each pixel with a Rayleigh prior
\begin{align}
    p(A_\mathrm{CMB}^\prime)
    &=
    \mathrm{Rayleigh}(A_\mathrm{CMB}^\prime,\sigma_\mathrm{CMB}^\prime)\\
    &=
    \frac{A_\mathrm{CMB}^\prime}{{\sigma_\mathrm{CMB}^\prime}^2}\mathrm{e}^{-{A_\mathrm{CMB}^\prime}^2/2{\sigma_\mathrm{CMB}^\prime}^2},
\end{align}
where $A_\mathrm{CMB}^\prime$ is the amplitude of the CMB fluctuation (in thermodynamic units) and $\sigma_\mathrm{CMB}^\prime=0.27,\mu\mathrm{K}_\mathrm{thermo.}$ is the standard deviation of the CMB polarization fluctuations (again in thermodynamic units).
\footnote{The quadrature sum of two Gaussian random variables with zero mean and common variance is a new random variable with a Rayleigh distribution.}

\begin{figure}
    \centering
    \includegraphics[width=\columnwidth]{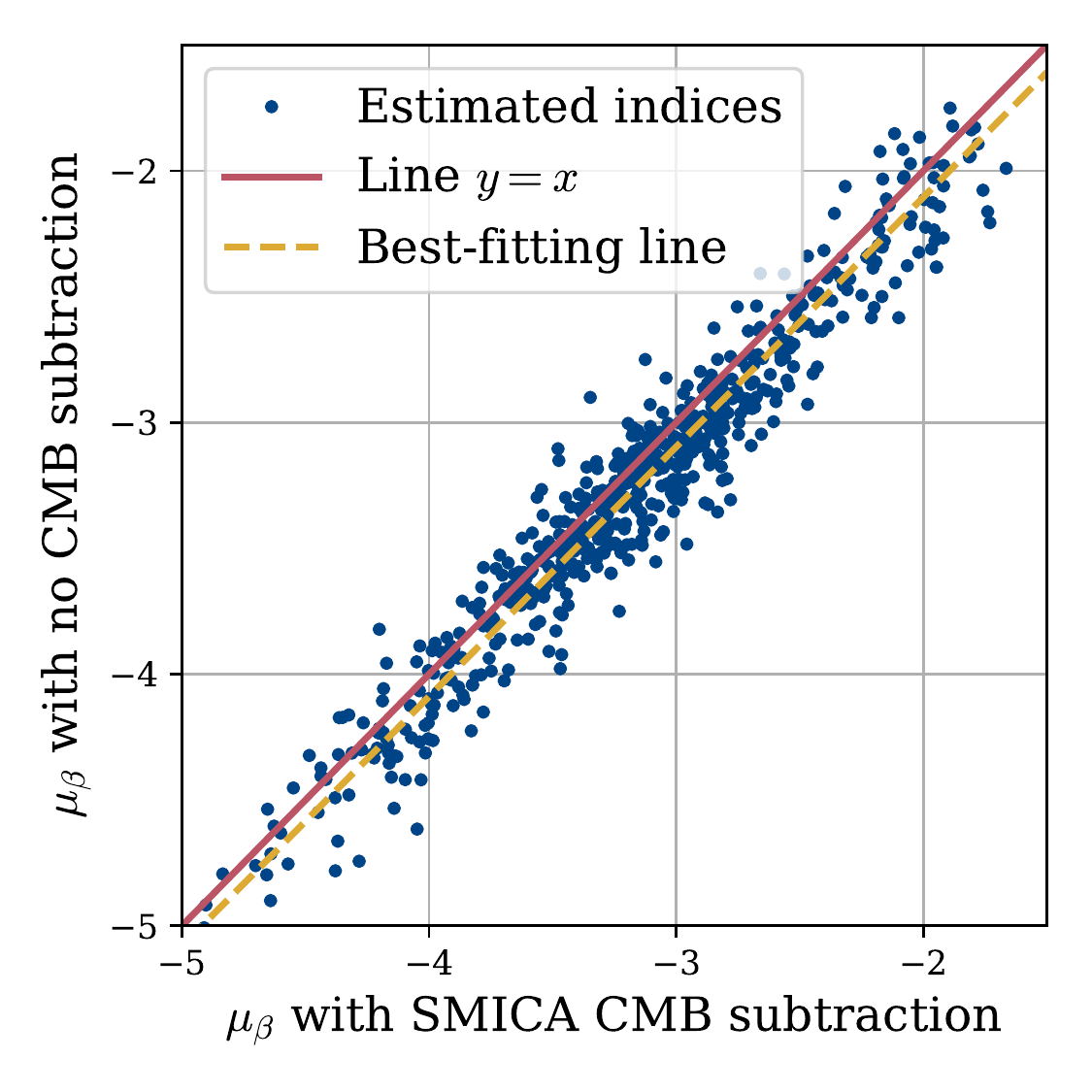}
    \caption{Scatter plot of the spectral indices in pixels identified by the clustering algorithm when the SMICA estimate of the CMB is subtracted ($x$-axis) and when instead marginalizing over the CMB amplitude ($y$-axis).
    The line $y=x$ is shown in \textit{solid red}.
    The best-fitting line is shown in \textit{dashed yellow}) and has equation $y=0.99x-0.12$.}
    \label{fig:cmbSubCompareScatter}
\end{figure}

Of all the tests we made to the robustness of our estimated spectral index map, this had the largest impact on the estimated spectral indices.
A scatter plot of the estimated spectral indices when subtracting the SMICA CMB estimate to the indices when instead marginalizing over the CMB amplitude is shown in Figure~\ref{fig:cmbSubCompareScatter}.
The points lie along the line $y=0.99x-0.12$, i.e. a line with slope very close to unity but a non-zero offset.
By marginalizing over the CMB amplitude we shift the spectral index posterior distributions to steeper values and also widen them by around 13\,per cent.

None the less, we still find significant differences between the average spectral index in the region of the north and south fermi bubbles.
The mean spectral indices in the bubbles and the Bayes factor that quantifies the evidence for the means being different, calculated
without the SMICA CMB estimate subtracted and instead marginalizing over the CMB amplitude, are listed in the first column of Table~\ref{tab:lobes_sigma_P_scaling}.
Subtracting the SMICA CMB estimate shifts the averages to steeper values and reduces the Bayes factor to $4\times10^5$, which is still decisive evidence that the averages are different.

\subsubsection{$I$ to $P$ leakage in the 44\,GHz map}
$I$ to $P$ leakage in the \textit{Planck}~44\,GHz map would result in shallower spectral indices, particularly in the north where the emission is brighter.
There are two reasons we believe this to be an unlikely cause;
firstly the bandpass mismatch $a$-factors for the 44\,GHz horns are less than those for the 30\,GHz horns \citep{PlanckCollaboration2018b}, secondly
typical foreground spectra are falling over this frequency range and so would also require the total intensity foreground emission in this region to be anomalously bright at 44\,GHz compared to 30\,GHz.

\subsubsection{Depolarization in the 30\,GHz map}
Depolarization, caused by Faraday rotation in the \textit{Planck}~30\,GHz survey, could lead to the measurement of spuriously shallow synchrotron spectral indices. Faraday rotation effects are expected to be more significant close to the Galactic plane, however these same effects are expected to be small at frequencies $\sim 30\,\mathrm{GHz}$. In the worst affected regions directly in the Galactic plane, we may see Faraday depths $\phi \approx 500\,\mathrm{rad/m^{2}}$ \citep{Wolleben2019,Oppermann2012}. Assuming the polarization angle varies linearly with the square of the wavelength, $\lambda^2$, we can estimate the worst Faraday rotation effects to induce position angle rotations $\Delta\psi\sim 3^{\circ}$. It is worth noting here that, in reality the position angle dependence for synchrotron emission will not vary linearly with $\lambda^2$. However this estimate is at least indicative of the low level of expected depolarization even in the most affected parts of the sky. Further, observations of the Faraday depth across the sky show that the level of Faraday rotation is even lower over the full extent of the Fermi bubbles. For these reasons it seems unlikely that the consistent estimation of shallow spectral indices in the northern bubble is the result of depolarization at $30\,\mathrm{GHz}$.

\subsubsection{Zero level of prior}
A systematic offset in the expected polarized intensity could result in shallower spectral indices when the signal-to-noise in the 44\,GHz map is less than that in the 30\,GHz map.
To test the impact of this we subtracted an 8.9\,K offset from the Haslam total intensity map that formed the prior.
This is the monopole found in the Haslam map by \citet{Wehus2014} using $T-T$ plots.

The average spectral indices are hardly affected by this change and actually the evidence for the average in the bubbles being different increases by around a factor of two.

\subsubsection{Astrophysical source}
Finally, the measured spectral indices may in fact be caused by some astrophysical source.
The hard-spectrum emission in the northern bubble is unlikely to be polarized free-free emission.
Firstly, free-free emission is typically observed to have a spectral index of $\beta\approx-2.1$, whilst we find an average spectral index in the northern bubble of $\beta=-2.36\pm 0.09$. Secondly, free-free emission is expected to be intrinsically un-polarized due to the random scattering directions giving rise to the emission. 
Free-free emission can be polarized at the edges of optically dense \ion{H}{ii} clouds, where the polarization fraction could reach $10\%$ \citep{Rybicki1985,Keating1998}. However, generally the polarization fraction will be close to zero.

In previous analyses with \textit{WMAP} and \textit{Planck} total intensity data, it was argued that the Galactic Haze emission likely arose from a hard synchrotron component.
In our analysis we find the polarized spectral indices in the southern bubble to be similar to the soft synchrotron emission that dominates over most of the sky.
The spectral indices in the northern bubble are harder.
It is important to note that these regions have many overlapping features and along the line-of-sight these features can sum to give unusual spectra, particularly in polarization \citep[for example][]{Lazarian2016}.
However, we have significant detections of shallower spectral indices in the region defined by the northern \textit{Fermi} bubble.
\section{Conclusions} \label{sec:conclusions}
We have used Bayesian methods to estimate the spectral index between the \textit{Planck} 30 and 44\,GHz polarized intensity maps.
We modelled the polarized intensities as Rician random variables and used a prior formed from the multiplication of the objective reference prior for the problem with a maximum entropy prior that incorporates total intensity information.
We tested our method on simulated data and found it gave unbiased estimates with uncertainties representative of the scatter on the data.

In each pixel we calculated the mean, standard deviation and Kullback-Liebler divergence from the prior of the spectral index posterior distribution.
We used a clustering algorithm on these parameters to identify pixels with good detections of the spectral index, and 
found good detections in 747 pixels (which corresponds to approximately one quarter of the sky).

We fitted for the mean and intrinsic scatter of the spectral index, assuming that the true spectral index across the sky is a Gaussian random variable and that the posterior distributions are Gaussian.
The global average spectral index that we found both across the whole sky and in pixels identified by the clustering algorithm are consistent with values found by others (these are summarised in Table~\ref{tab:real_polBetas}).
Across all pixels we found an average $\left< \beta\right>=-2.99\pm0.03$ and in pixels identified by the clustering algorithm we found an average of $\left<\beta\right>=-3.12\pm0.04$.
The angular power spectrum of our spectral index map is consistent with the spectrum found by \citet{Krachmalnicoff2018} (Figure~\ref{fig:results_beta_spectrum}).

We found a statistically significant difference between the average spectral index in the North and South \textit{Fermi} bubbles.
Only including pixels identified by the clustering algorithm,
the average spectral index in the Northern bubble is $\left<\beta\right>=-2.36\pm0.09$, the average spectral index in the Southern bubble is $\left<\beta\right>=-3.00\pm0.05$, and 
the average spectral index across both bubbles is $\left<\beta\right>=-2.72\pm0.06$.
The average across both bubbles including all pixels, not just those identified by the clustering algorithm, is $\left<\beta\right>=-2.57\pm0.08$ and this is consistent with the spectral index found by \citet{PlanckCollaboration2012a} who treated both bubbles as a single component and found a spectral index of $-2.56\pm0.05$.

The largest potential systematic error in this analysis is the CMB subtraction step.
Wen we marginalize over the CMB amplitude (instead of subtracting an estimate) the indices are typically shifted to more negative values, $\Delta\beta=-0.12$, and the uncertainties are increased by around 13\,per cent.

In future work, 
repeating this analysis with higher signal-to-noise maps will increase the fraction of the sky with detections and
the weakly informative part of the prior could be improved with a spatially varying temperature spectral index and polarization fraction.

\section*{Acknowledgements}
We would like to thank Clive Dickinson and Tim Pearson for comments on a draft of this paper.
We would also like to thank Clive Dickinson further for useful discussions about diffuse Galactic emission at radio frequencies.




\bibliographystyle{mnras}
\bibliography{library}








\bsp	
\label{lastpage}
\end{document}